\documentclass[aps,pre,epsf,superscriptaddress,amsmath,amssymb,amsfonts,twocolumn,showpacs]{revtex4-1}

\usepackage{graphicx}
\usepackage{epsfig}
\usepackage{dcolumn}
\usepackage{bm}
\usepackage{braket}
\usepackage{amsmath}
\usepackage{color}
\usepackage{hyperref}
\newcommand{\abs}[1]{\left| #1 \right|} 

\usepackage{multirow}
\DeclareMathOperator{\sech}{sech} 

\begin{document}

\title{Dissipative correlated dynamics of a moving impurity\\ immersed in a Bose-Einstein Condensate} 

\author{S. I. Mistakidis}
\affiliation{Center for Optical Quantum Technologies, Department of Physics, University of Hamburg, 
Luruper Chaussee 149, 22761 Hamburg Germany}
\author{F. Grusdt}
\affiliation{Department of Physics and Institute for Advanced Study, Technical University of Munich, 85748 Garching, Germany} 
\affiliation{Munich Center for Quantum Science and Technology (MCQST), Schellingstr. 4, D-80799 M\"unchen, Germany}
\author{G.M. Koutentakis}
\affiliation{Center for Optical Quantum Technologies, Department of Physics, University of Hamburg, 
Luruper Chaussee 149, 22761 Hamburg Germany} 
\author{P. Schmelcher}
\affiliation{Center for Optical Quantum Technologies, Department of Physics, University of Hamburg, 
Luruper Chaussee 149, 22761 Hamburg Germany} \affiliation{The Hamburg Centre for Ultrafast Imaging,
Universit\"{a}t Hamburg, Luruper Chaussee 149, 22761 Hamburg, Germany}

\date{\today}

\begin{abstract} 

We unravel the nonequilibrium correlated quantum quench dynamics of an impurity traveling through a harmonically confined Bose-Einstein 
condensate in one-dimension. 
For weak repulsive interspecies interactions the impurity oscillates within the bosonic gas. 
At strong repulsions and depending on its prequench position the impurity moves towards an edge of the bosonic medium and subsequently equilibrates. 
This equilibration being present independently of the initial velocity, the position and the mass of the impurity is inherently related 
to the generation of entanglement in the many-body system. 
Focusing on attractive interactions the impurity performs a damped oscillatory motion within the bosonic bath, a behavior that becomes more evident 
for stronger attractions. 
To elucidate our understanding of the dynamics an effective potential picture is constructed. 
The effective mass of the emergent quasiparticle is measured and found to be generically larger than the bare one, especially for strong attractions. 
In all cases, a transfer of energy from the impurity to the bosonic medium takes place. 
Finally, by averaging over a sample of simulated in-situ single-shot images we expose how the single-particle density distributions and the two-body 
interspecies correlations can be probed. 
 
\end{abstract}

\maketitle

\section{Introduction} 

Ultracold atoms offer an excellent platform to study highly imbalanced multicomponent systems, such as impurities 
immersed in a Bose-Einstein Condensate (BEC) or in a Fermi sea \cite{Scazza,Kohstall,Schirotzek}, due to their exquisite degree of controlability. 
Indeed the interaction between the impurities and host atoms is tunable via Feshbach resonances \cite{Chin,Kohler} and the emergent 
many-body states can be characterized e.g. with the aid of radiofrequency and Ramsey spectroscopy \cite{Koschorreck,Kohstall,Cetina,Cetina_interferometry} and 
in-situ measurements with single-site resolution \cite{Fukuhara,Sherson}. 
Mobile impurities interacting with a surrounding quantum many-body environment form quasiparticles, such as polarons \cite{Massignan,Schmidt_rev}, 
originally introduced by Landau \cite{Landau}, when the medium consists of neutral atoms that are not exposed to any external field. 
The dressing of the impurity atoms from the collective excitations of their host leads to alterations of their properties including their effective mass \cite{Grusdt_1D,Ardila_MC}, 
induced interactions \cite{induced_int_artem,Mistakidis_Fermi_pol}, formation of bound states e.g. bipolarons \cite{Bipolaron,Massignan,Schmidt_rev,Kevin}, as well as dramatic changes during their 
nonequilibrium dynamics \cite{dynamics_Artem,Mistakidis_orth_cat,Mistakidis_two_imp_ferm,Mistakidis_eff_mass,Grusdt_RG,Shchadilova,Kamar}. 
Another important feature is that the impurity subsystem constitutes a few-body setting evincing the inescapable necessity of taking correlation effects into account. 

The controllable immersion of single or multiple impurities into a many-body environment have recently led to the experimental observation of 
Bose \cite{Jorgensen,Hu,Catani1,Fukuhara,Yan_bose_polarons} and Fermi \cite{Scazza,Koschorreck,Kohstall} polarons. 
These experiments triggered an intense theoretical activity in order to describe the polaron characteristics by operating within different frameworks \cite{Grusdt_approaches,Rath_approaches}. 
These include, but are not restricted to, the mean-field approximation \cite{Astrakharchik,Cucchietti,Kalas,Bruderer1}, the Fr\"ohlich model \cite{Bruderer,Privitera,Casteels1,Casteels2,Kain,Tempere_path_int}, 
variational methods \cite{Mistakidis_eff_mass,Mistakidis_orth_cat,Mistakidis_Fermi_pol,Jorgensen,Ardila_MC}, effective Hamiltonian 
approaches \cite{Effect_hamilt,Effect_hamilt1,induced_int_artem,dynamics_Artem} and renormalization group techniques \cite{Grusdt_RG,Grusdt_strong_coupl,Grusdt_approaches}.  
While the majority of these investigations have been mainly focused on the equilibrium properties of the emergent quasiparticles, the dynamics 
of impurities is far less explored. 
In this context notable examples include the observation of self-trapping phenomena \cite{Cucchietti_self_trap,Schecter_self_trap}, 
orthogonality catastrophe events \cite{Mistakidis_orth_cat}, generation of dark-bright solitons \cite{Grusdt_1D,Mistakidis_two_imp_ferm}, 
transport properties of impurities in optical lattices \cite{Cai_transp,Johnson_transp} as well as collisional aspects \cite{few_body_col,Lausch_col,Lausch_col1} of an impurity injected 
into a gas of Tonks-Girardeau 
bosons \cite{Burovski_col,Lychkovskiy_col,Lychkovskiy_col1,Lychkovskiy_col2,Meinert,Flutter,Flutter1,Gamayun_col}. 

In this latter context, Bloch-oscillations of impurities in the absence of a lattice \cite{Meinert}, 
long-lived oscillations \cite{Flutter,Flutter1}, as well as relaxation of moving impurities \cite{Burovski_col,Lychkovskiy_col,Lychkovskiy_col1,Lychkovskiy_col2} 
have been observed in one-dimension. 
The majority of these investigations have been focusing on a Tonk-Girardeau gas of host atoms in homogeneous space. 
Yet the collisional dynamics of an impurity particle penetrating a weakly or intermediate repulsively interacting quantum bosonic gas 
being harmonically trapped can be much more involved. 
Indeed, in this case the dynamics of the impurity might exhibit a completely different behavior compared to the aforementioned settings for the following reasons. 
First the finite interactions between the host atoms will give rise to fundamentally different scattering properties between the impurity and the bosonic medium. 
In this sense it would be particularly interesting to examine the dynamical response of the impurity for different interspecies repulsive and attractive interactions and 
study how the response regimes depend on the velocity (subsonic, sonic and supersonic) of the impurity. 
Note also that the initial velocity of the impurity is expected to give rise to a much more involved dynamics compared to the zero 
velocity case since it will trigger multiple scattering events between the impurity and the BEC. 
Concordantly, one can e.g. infer whether long-lived oscillations occur \cite{Flutter,Flutter1,Meinert} for subsonic or sonic impurities.  
Moreover, the presence of the external harmonic trap confines the bosonic bath to a finite spatial region and it would be important to inspect under 
what conditions the impurity can escape the BEC. 
Another intriguing prospect is to examine if the impurity forms a strongly correlated (entangled) state with the bosonic bath generating a quasiparticle and unveil 
the correlation effects of its dynamics \cite{Mistakidis_orth_cat,Flutter}. 
Certainly the properties of the generated quasiparticle such as its effective mass are of significant importance.  

To address these inquiries in the present work we investigate the interspecies interaction quench dynamics of a moving impurity initially modeled by a coherent state which 
penetrates a repulsively interacting and harmonically trapped bosonic gas. 
To simulate the correlated quantum dynamics of the mixture we invoke the Multi-Layer Multi-Configuration Time-Dependent Hartree 
Method for Atomic Mixtures (ML-MCTDHX) \cite{MLX,MLB1}, which is a non-perturbative variational method capturing all interparticle correlations. 
We find that the dynamics of the impurity exhibits different response regimes depending on the value of the postquench interspecies 
interaction strength \cite{Mistakidis_orth_cat,Mistakidis_two_imp_ferm,Grusdt_1D}. 
In particular, for weak postquench interspecies repulsive interactions the subsonic impurity undergoes a dipole motion with a larger frequency for 
increasing coupling. 
This is in sharp contrast to the behavior of an initially stationary impurity which, following an interspecies interaction quench, performs a breathing motion 
inside the BEC as analyzed in Ref. \cite{Mistakidis_orth_cat}. 
Strikingly enough, at strong interspecies interactions which exceed the bosonic intraspecies ones the impurity 
moves towards the edge of the BEC background and thereafter equilibrates around its Thomas-Fermi radius. 
This behavior of the moving impurity observed for strong repulsions takes place independently of its characteristics e.g. initial velocity, 
prequench position, trapping frequency and mass and occurs due to the presence of correlations. 
Importantly, the density of the impurity approaches selectively the smaller distant edge of the Thomas-Fermi radius with respect to its prequench 
position. 
This result is altered for a zero velocity impurity whose density at such strong repulsions breaks into two fragments which exhibit a dissipative oscillatory 
motion at the edges of the Thomas-Fermi profile of the bosonic gas, e.g. see Ref. \cite{Mistakidis_orth_cat}. 
In all cases, the bosonic bath shows weak distortions from its initial Thomas-Fermi profile and shallow density dips 
built upon the bosonic density thus imprinting the motion of the impurity. 
Indeed, the response of the bosonic medium for a zero velocity impurity undergoes weak amplitude collective breathing 
oscillations as it has been demonstrated in Ref. \cite{Mistakidis_orth_cat}. 

Referring to quenches towards attractive interspecies interactions we showcase that the impurity performs a damped oscillatory motion within the 
bosonic bath, a behavior that becomes more evident for stronger attractions. 
As a result the BEC develops a density peak at the location of the impurity. 
An effective potential picture is also developed for each case in order to provide an intuitive understanding of the resulting dynamics 
of the impurity \cite{Mistakidis_orth_cat,Grusdt_1D}. 
Note that this effective potential is greatly affected by the motion of the impurity, causing significant deformations in the Thomas-Fermi profile 
of the bosonic medium, a result that is in contrast to the initially stationary impurity case. 
Inspecting the individual energy contributions of each species we reveal that the impurity dissipates energy into the bosonic medium \cite{Mistakidis_orth_cat,Nielsen,Lampo}, 
a phenomenon that is more enhanced for stronger interactions of either sign. 
Employing the Von-Neumann entropy we unveil the presence of interspecies correlations in the course of the evolution. 
It is worth mentioning that energy exchange processes and the development of correlations between the impurity and the BEC are generic 
phenomena appearing in impurity physics and their emergence is almost independent of the considered quench scenario. 
Moreover, we estimate the effective mass of the emergent quasiparticle showcasing that for attractive interactions it is larger than the bare one tending 
to the latter when approaching the non-interacting limit and becoming slightly larger to its bare value for repulsive interactions. 
Finally, we provide experimental links of our findings by simulating single-shot absorption measurements. 
We demonstrate that by averaging a sample of in-situ 
images the quench-induced correlated dynamics can be adequately reproduced on the single-particle density level \cite{mistakidis_phase_sep,Erdmann_phase_sep,Katsimiga_diss_flow}. 
Also by utilizing the simulated images on the co-moving frame of the impurity we showcase its imprint on the bosonic gas and discuss how 
the resulting imaging process probes the two-body interspecies correlations \cite{Koepsell}. 

This work is organized as follows. 
Section \ref{sec:theory} presents our setup, the employed many-body treatment and different observables 
that are used for the characterization of the dynamics. 
Subsequently, the resulting interspecies interaction quench dynamics towards repulsive [Sec. \ref{sec:quench_repulsive}] and attractive [Sec. \ref{sec:quench_attractive}] 
interactions is discussed. 
The effective mass of the emergent quasiparticle is analyzed in Section \ref{sec:effective_mass}, while in Section \ref{sec:single_shots} we present 
the simulation of in-situ single-shot images. 
We summarize and discuss future perspectives in Section \ref{sec:conclusions}. 
In Appendix \ref{sec:single_shot_algorithm} we elaborate on the numerical implementation of the single-shot process and in Appendix \ref{sec:convergence_numerics} 
we demonstrate the convergence of our many-body calculations.

\section{Theoretical Framework}\label{sec:theory}

\subsection{Setup and Quench Protocol}\label{sec:hamiltonian}

We consider a highly particle imbalanced bosonic mixture consisting of a single impurity atom $N_I=1$ and $N_B=100$ bosons constituting the majority species (bath). 
The many-body Hamiltonian of the system consisting of mass balanced species, i.e. $m_A=m_B\equiv m$, which are trapped in the same external one-dimensional 
harmonic oscillator potential of frequency $\omega_A=\omega_B=\omega$ reads  
\begin{equation}
\begin{split}
	\label{eq:hamiltonian}
	H = \sum_{\sigma = B,I} \sum_{i = 1}^{N_{\sigma}}\left[-\frac{\hbar^2}{2m}\left(\frac{\mathrm{d}}{\mathrm{d}x_i^{\sigma}}\right)^2 + 
	\frac{1}{2} m \omega (x_i^{\sigma})^2\right] \\+ g_{BB} \sum_{i<j} \delta(x_i^B - x_j^B) + g_{BI} \sum_{j = 1}^{N_{B}} \delta(x_j^B-x_1^I).
\end{split}
\end{equation}
Within the $s$-wave scattering limit which is the dominant interaction process in the ultracold regime both the intra- and interspecies 
interactions are modeled by a contact potential with effective coupling constants $g_{BB}$ and $g_{BI}$. 
More specifically, the effective one-dimensional coupling strength \cite{Olshanii} is given by  
${g_{\sigma \sigma'}} =\frac{{2{\hbar ^2}{a^s_{\sigma \sigma'}}}}{{\mu a_ \bot ^2}}{\left( {1 - {\left|{\zeta (1/2)} \right|{a^s_{\sigma \sigma'}}}
/{{\sqrt 2 {a_ \bot }}}} \right)^{ -1}}$, where $\sigma,\sigma'=B, I$ and $\mu=\frac{m}{2}$ refers to the corresponding reduced mass. 
Here, ${a_\bot } = \sqrt{\hbar /{\mu{\omega _ \bot }}}$ is the transversal length scale characterized by a transversal confinement frequency ${{\omega _ \bot }}$.  
Also, ${a^s_{\sigma \sigma'}}$ is the three-dimensional $s$-wave scattering length within ($\sigma=\sigma'$) or between ($\sigma \neq \sigma'$) 
the two species. 
Experimentally $g_{\sigma\sigma'}$ can be adjusted through ${a^s_{\sigma \sigma'}}$ via Feshbach resonances \cite{Kohler,Chin} 
as well as by manipulating ${{\omega _ \bot }}$ by means of confinement-induced resonances \cite{Olshanii}. 
Throughout this work we use a trapping frequency $\omega=0.1\approx2\pi\times 20~ {\rm Hz}$, unless it is stated otherwise. 
To restrict the dynamics to one dimension one can e.g. assume the experimentally relevant transversal confinement 
$\omega_{\perp}\approx 2\pi \times 200~ {\rm Hz}$ which is typical for one-dimensional experiments \cite{Bersano,Jochim_resolved}. 
Additionally, the intraspecies interactions are kept fixed to $g_{BB}=1.0$ while the interspecies one, $g_{BI}$, varies upon considering a quench taking 
values in the repulsive or the attractive regime. 
Due to the above-mentioned assumptions our system can be well approximated by a binary BEC of $^{87}$Rb atoms prepared in the hyperfine 
states $\Ket{F=1, m_F=-1}$ and $\Ket{F=2, m_F=1}$ \cite{Egorov}.  

For convenience, in the following, the many-body Hamiltonian of Eq. (\ref{eq:hamiltonian}) is rescaled in units of $\hbar  \omega_{\perp}$. 
Then the corresponding length, time, and interaction strengths are provided in terms of
$\sqrt{\frac{\hbar}{m \omega_{\perp}}}$, $\omega_{\perp}^{-1}$ and 
$\sqrt{\frac{\hbar^3 \omega_{\perp}}{m}}$ respectively. 
Also, the spatial extension of our system is limited by employing hard-wall boundary conditions at $x_\pm=\pm80$. 
The location of the latter does not affect the dynamics since we do not observe any significant density population beyond $x_\pm=\pm40$.  

\begin{figure}[ht]
 	\centering
  	\includegraphics[width=0.45\textwidth]{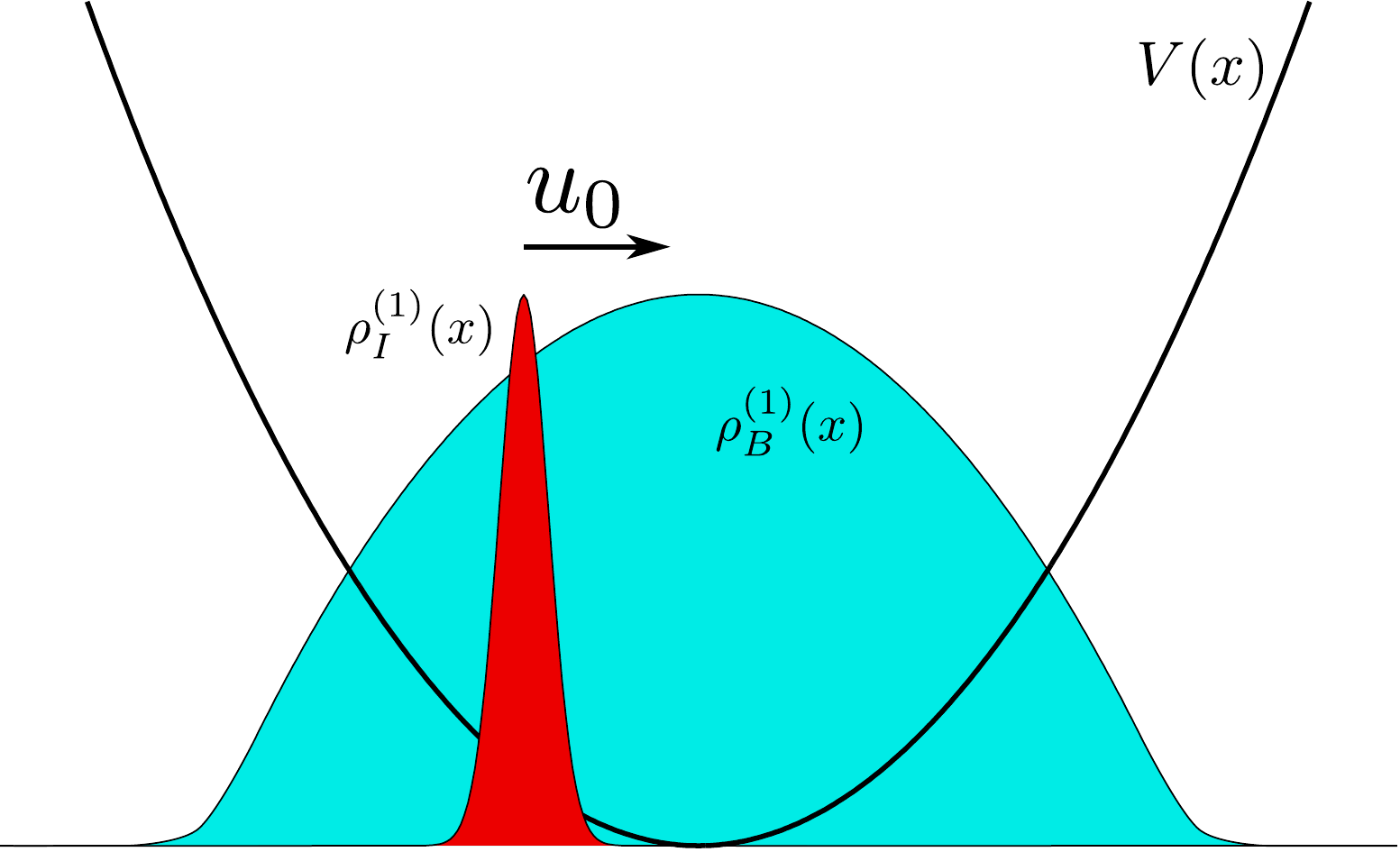}
     \caption{Sketch of the dynamical quench protocol. 
     An impurity modeled initially by a coherent state travels through a harmonically trapped BEC medium with an initial velocity $u_0$. 
     At $t=0$ an interspecies interaction quench is performed from $g_{BI}=0$ to a finite positive or negative value of $g_{BI}$.} 
     \label{fig:1}
\end{figure} 

In order to examine the correlation effects arising due to the injection of the impurity into the bosonic gas we follow the protocol outlined below. 
The bosonic medium is initially prepared into its ground state for $g_{BB}=1$. 
The impurity resides in a coherent state characterized by an initial velocity $u_0$ and it is instantaneously localized around $x_0$. 
In particular, its wavefunction assumes the form 
\begin{equation}
 \Psi^I(x^I;t=0)=\bigg(\frac{m\omega}{\pi \hbar}\bigg)^{1/4} e^{-\frac{m\omega}{2\hbar}(x^I-x_0)^2+ik_0(x^I-x_0)}, \label{coherent_state}
\end{equation}
where $k_0$ is its initial wavenumber and $u_0=k_0/m$ its initial velocity. 
To trigger the collisional dynamics between the impurity and the bosonic medium we suddenly switch on at $t=0$ the interspecies interaction 
strength to repulsive [Section \ref{sec:quench_repulsive}] or attractive [Section \ref{sec:quench_attractive}] interactions and monitor 
the time-evolution of the system, see Fig. \ref{fig:1} for a sketch. 
Experimentally our protocol can be realized as follows. 
A magnetic potential gradient that shifts the minimum of the external trap of the impurity with respect to the bath by $\delta x=u_0/\omega$ is first employed \cite{Du_gradient}. 
Then, in order to import an initial momentum onto the impurity this gradient is swiched off letting the impurity to evolve and a 
Feshbach resonance is utilized to perform the interaction quench when the impurity is at position $x_0$.

\subsection{Many-Body Treatment}\label{ML_ansatz}

To obtain the ground state and most importantly to explore the quench dynamics of the bosonic mixture we numerically solve the underlying 
many-body Schr{\"o}dinger equation by employing the ML-MCTDHX \cite{MLX,MLB1} method. 
It is an ab-initio method which rests on expanding the systems' many-body wavefunction in terms of a time-dependent and variationally optimized 
basis enabling us to take into account both the inter- and the intraspecies correlations of the binary system. 

To include interspecies correlations into our many-body wavefunction ansatz we introduce $k=1,2,\dots,D$ different species functions for each component, 
namely $\Psi^{B}_k (\vec x^{B};t)$ and $\Psi^{I}_k (x^{I};t)$ respectively. 
Here, $\vec x^{B}=\left( x^{B}_1, \dots, x^{B}_{N_{B}} \right)$ and $x^{I}$ denote the spatial coordinates of each species with 
$N_B$ and $N_I\equiv1$ being the number of bath and impurity atoms, respectively. 
Then the many-body wavefunction, $\Psi_{MB}$, is expressed in the form of a truncated Schmidt decomposition \cite{Horodecki} of rank $D$ as follows  
\begin{equation}
\Psi_{MB}(\vec x^B, x^I;t) = \sum_{k=1}^D \sqrt{ \lambda_k(t) }~ \Psi^B_k (\vec x^B;t) \Psi^I_k ( x^I;t).    
\label{Eq:WF}
\end{equation} 
The Schmidt coefficients $\lambda_k(t)$ provide a measure of the entanglement between the two species (see also below) and will be 
referred to, in the following, as the natural species populations of the $k$-th species function. 
Indeed, the system is termed entangled \cite{Roncaglia} or interspecies correlated if at least two different $\lambda_k(t)$ possess a nonzero value 
since in this case $\Psi_{MB}$ is not a direct product of two states. 

Moreover, in order to account for the intraspecies correlations of the system we further express each of the above-mentioned species functions 
$\Psi^{B}_k (\vec x^{B};t)$ with respect to permanents consisting of $d_{B}$ distinct time-dependent 
single-particle functions (SPFs) $\varphi_1^B,\dots,\varphi_{d_{B}}^B$. 
As a result, $\Psi^{B}_k (\vec x^{B};t)$ of the bosonic gas reads 
\begin{equation}
\begin{split}
&\Psi_k^{B}(\vec x^{B};t) = \sum_{\substack{l_1,\dots,l_{d_{B}} \\ \sum l_i=N_B}} c_{k,(l_1,
\dots,l_{d_{B}})}^B(t) \times \\ & \sum_{i=1}^{N_{B}!} \mathcal{P}_i
 \left[ \prod_{j=1}^{l_1} \varphi_1^B(x_j;t) \cdots \prod_{j=1}^{l_{d_{B}}} \varphi_{d_{B}}^B(x_{K(d_{B})+j};t) \right],   
 \label{Eq:SPF_bath}
 \end{split}
\end{equation} 
where $\mathcal{P}$ is the permutation operator which exchanges the particle positions $x_{\nu}^{B}$, $\nu=1,\dots,N_{B}$ within the SPFs.  
Also, $K(r)\equiv \sum_{\nu=1}^{r-1}l_{\nu}$ with $l_{\nu}$ being the occupation number of the $\nu$th SPF and $r\in\{1,2,\dots,d_{B}\}$, while 
$c_{k,(l_1,\dots,l_{d_{B}})}^B(t)$ are the time-dependent expansion coefficients of a certain permanent. 
Correspondingly, the species functions $\Psi^{I}_k (x^{I};t)$ of the impurity are expressed as follows 
\begin{equation}
\begin{split}
&\Psi_k^{I}( x^{I};t) = \sum_{p=1}^{d_I} c_{k,(l_1=0,\dots,l_p=1,\dots,l_{d_{B}}=0)}^I(t) \varphi_{p}^I(x^I;t),  
 \label{Eq:SPF_impurity}
 \end{split}
\end{equation} 
with $c_{k,(l_1=0,\dots,l_p=1,\dots,l_{d_{B}}=0)}^I(t)$ being the respective time-dependent expansion 
coefficients on the SPFs $\varphi_1^I,\dots,\varphi_{d_{I}}^I$ of the species $I$.  

To solve the underlying Schr{\"o}dinger equation we need to determine the corresponding ML-MCTDHX equations of motion \cite{MLX,Kohler_fabian}. 
The latter can be accomplished by following e.g. the Dirac-Frenkel variational principle \cite{Frenkel,Dirac} for the 
many-body ansatz given by Eqs.~(\ref{Eq:WF}), (\ref{Eq:SPF_bath}) and (\ref{Eq:SPF_impurity}). 
This way we arrive at $D^2$ linear differential equations of motion for the coefficients $\lambda_k(t)$ which are coupled to 
a set of $D[$ ${N_B+d_B-1}\choose{d_B-1}$+$d_I$] non-linear integro-differential equations for the species functions and $d_B+d_I$ 
integro-differential equations for the SPFs. 
Finally, let us remark that within ML-MCTDHX it is also possible to operate at different orders of approximation. 
For instance, we can retrieve the corresponding mean-field wavefunction \cite{Pethick_book} of the bosonic mixture in the limit of $D=d_B=d_I=1$. 
Namely 
\begin{equation}
\begin{split}
&\Psi_{MF}(\vec x^{B}, x^{I};t) =
	\frac{1}{\sqrt{N_B!}}\prod_{j=1}^{N_B}\varphi_1^B(x_j^{B};t)\varphi_1^I(x^{I};t).  
\label{Eq:MF}
\end{split}
\end{equation} 
Recall that in this approximation both intra- and interspecies correlations are neglected since the system is described by one single-particle function 
for each of the species \cite{Pethick_book,mistakidis_phase_sep}.  

\subsection{Obervables of interest}\label{observables} 

To monitor the dynamics of each species after the quench we employ as a spatially resolved measure the $\sigma$-species one-body reduced density matrix 
\begin{equation}
\rho_\sigma^{(1)}(x,x^\prime;t)=\langle\Psi_{MB}(t)|\hat{\Psi}^{\sigma\dagger}(x)\hat{\Psi}^\sigma(x^\prime)|\Psi_{MB}(t)\rangle. 
\end{equation}
In this expression, $\hat{\Psi}^{\sigma}(x)$ [$\hat{\Psi}^{\sigma \dagger}(x)$] denotes the bosonic field operator that annihilates [creates] a $\sigma$-species 
boson at position $x$, satisfying also the standard bosonic commutation relations \cite{Pethick_book}. 
To be more specific, in the following, for simplicity we will resort to the corresponding $\sigma$-species one-body density defined as 
$\rho_\sigma^{(1)}(x;t)\equiv\rho_\sigma^{(1)}(x,x^\prime=x;t)$. 
Note also that the eigenfunctions and eigenvalues of $\rho_\sigma^{(1)}(x,x^\prime;t)$ are the so-called natural orbitals $\phi^{\sigma}_i(x;t)$ 
and natural populations $n^{\sigma}_i(t)$ \cite{mistakidis_phase_sep,MLX}. 
The natural orbitals are related with the SPFs [see Eqs. (\ref{Eq:SPF_bath}), (\ref{Eq:SPF_impurity})] via a time-dependent unitary transformation that diagonalizes 
the matrix $\rho_{\sigma;ij}^{(1)}=\int dx dx' \varphi_i^*(x,t)\varphi_j(x',t)\rho_{\sigma}^{(1)}(x,x';t)$, more details can be found in \cite{MLX,MLB1}. 
Moreover, each bosonic subsystem is said to be intraspecies correlated if more than one natural population possesses a macroscopic occupation, 
otherwise the corresponding subsystem is fully coherent. 
Indeed it can be easily shown that when $n_1^{\sigma}(t)=1$, $n_{i>1}^{\sigma}(t)=0$ holds, then $\Psi_{MB}(\vec x^B, x^I;t) \to \Psi_{MF}(\vec x^B, x^I;t)$, 
see also Eq.~(\ref{Eq:WF}) and Eq.~(\ref{Eq:MF}). 
As a result the degree of the $\sigma$ subsystem intraspecies correlations, and therefore the deviation of the many-body state from the mean-field one, 
can be theoretically quantified via $\lambda_{\sigma}(t) =1 -n_1^{\sigma}(t)$. 

To unveil the degree of interspecies correlations or entanglement during the nonequilibrium dynamics of the bosonic mixture we measure the so-called 
von-Neumann entropy \cite{Erdmann_phase_sep,Catani}. 
Recall that [see also the discussion in Section \ref{ML_ansatz} and Eq. \ref{Eq:WF})] the presence of interspecies correlations or entanglement can be inferred 
by the values of the higher than the first Schmidt coefficients, i.e. $\lambda_k(t)$ with $k>1$. 
The Schmidt coefficients $\lambda_k(t)$ are the eigenvalues of the species reduced density matrix e.g. 
$\rho^{N_{B}} (\vec{x}'^{B};t)=\int d x^{I} \Psi^*_{MB}(\vec{x}^{B}, 
x^{I};t) \Psi_{MB}(\vec{x}'^{B}, x^{I};t)$, with $\vec{x}^{B}=(x^{B}_1, \cdots, 
x^{B}_{N_{B-1}})$ [see also Eq. (\ref{Eq:WF})]. 
They are chosen to be ordered in a monotonically decreasing manner i.e. $\lambda_k>\lambda_{k+1}$, $k=1,2,\dots,D-1$. 
As a consequence, the system is species entangled or interspecies correlated when more than a single eigenvalue of $\rho^{N_{B}}$ 
are macroscopically populated, otherwise it is non-entangled. 
The von-Neumann entropy \cite{Erdmann_phase_sep,Catani,Mistakidis_two_imp_ferm} reads  
\begin{align}
S_{VN}(t)=-\sum\limits_{k=1}^D \lambda_{k}(t)\ln[\lambda_k(t)].\label{eq:entropy} 
\end{align} 
It can be easily deduced that in the mean-field limit $S_{VN}(t)=0$ since $\lambda_1(t)=1$, while for a many-body state where more than 
one Schmidt coefficients $\lambda_k$ are populated it holds that $S_{VN}(t)\neq0$. 

To track the position of the impurity in the course of the evolution we rely on its spatially averaged mean position. 
This enables us to assess the trajectory of the impurity given by 
\begin{equation}
 \braket{X_I(t)}=\braket{\Psi_{MB}(t)|\hat{x}^I|\Psi_{MB}(t)},\label{mean_position} 
\end{equation} 
where the one-body operator $\hat{x}^I=\int_{\mathcal{D}} dx x^I \hat{\Psi}^{\sigma \dagger}(x) \hat{\Psi}^{\sigma}(x)$ with $\mathcal{D}$ being the 
spatial extension of the impurity. 
Experimentally, $\braket{X_I(t)}$ can be measured via spin-resolved single-shot absorption images \cite{Jochim_resolved}. 
In particular, each image provides an estimate of the impurity position and $\braket{X_I(t)}$ can be obtained by averaging over a sample of such images.

\begin{figure*}[ht]
 	\centering
  	\includegraphics[width=0.9\textwidth]{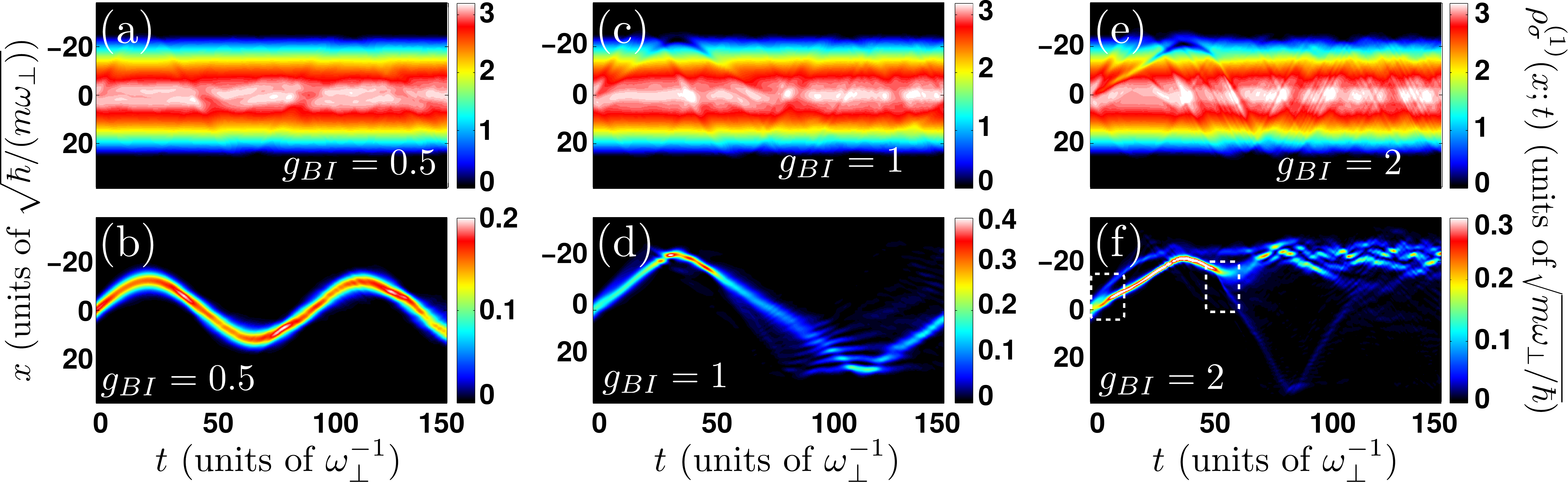}
     \caption{Time-evolution of the one-body density of the bath (upper panels) and the impurity (lower panels) for different interspecies repulsive interaction strengths 
     $g_{BI}$ (see legends). 
     The system consists of $N_{B}=100$ bosons initialized in their ground state with $g_{BB}=1$ and $N_I=1$ impurity atoms residing in a coherent state. 
     The latter is located at $x_0=0$ and it possesses an initial velocity $u_0=-u_c/2$ with $u_c$ denoting the speed of sound of the bosonic gas. 
     Both species are trapped in an external harmonic oscillator of frequency $\omega=0.1$. 
     To trigger the dynamics at $t=0$ we switch on the interspecies repulsion from $g_{BI}=0$ to a finite value (see legends). 
     The dashed rectangles in (f) mark density emission events of the impurity. } 
     \label{fig:2}
\end{figure*} 

Concluding, we remark that our predictions can be directly tested in state-of-the-art experimental 
settings \cite{Yan_bose_polarons,Jorgensen,Hohmann_Rb_Cs,Spethmann_Rb_Cs,Fukuhara,Meinert,Catani,Mayer_doping}. 
Indeed, the initial state of the impurity is prepared by utilizing a magnetic gradient, while  
the employed quench protocol can be realized with the aid of a Feshbach resonance. 
Moreover, the main quantities used to monitor the dynamics such as the single-particle density and the trajectory of the impurity can be 
experimentally tracked via in-situ single-shot absorption measurements as we discuss in Section \ref{sec:single_shots}.

\section{Quench Dynamics Towards Repulsive Interactions}\label{sec:quench_repulsive} 

In the following we investigate the collisional dynamics of a moving single impurity, $N_I=1$, inside a harmonically trapped many-body bosonic 
bath of $N_B=100$ atoms following an interspecies interaction quench to repulsive interactions. 
The many-body bath is initialized in its ground state with $g_{BB}=1$ exhibiting a Thomas-Fermi profile of radius $R_{TF}\approx25$. 
On the other hand, the impurity is non-interacting with the bath, $g_{BI}=0$, and resides in a 
coherent state [see Eq. (\ref{coherent_state})]. 
Its initial velocity is $u_0=-u_c/2$, with $u_c$ being the speed of sound of the BEC background, and therefore it is subsonic.  
The dynamics is triggered by performing an interspecies interaction quench to positive $g_{BI}$ values at $t=0$ where the impurity is 
located at position $x_0=0$.

\subsection{Single-Particle Density Evolution and Effective Potential}\label{sec:density_repulsive}

\begin{figure}[ht]
 	\centering
  	\includegraphics[width=0.45\textwidth]{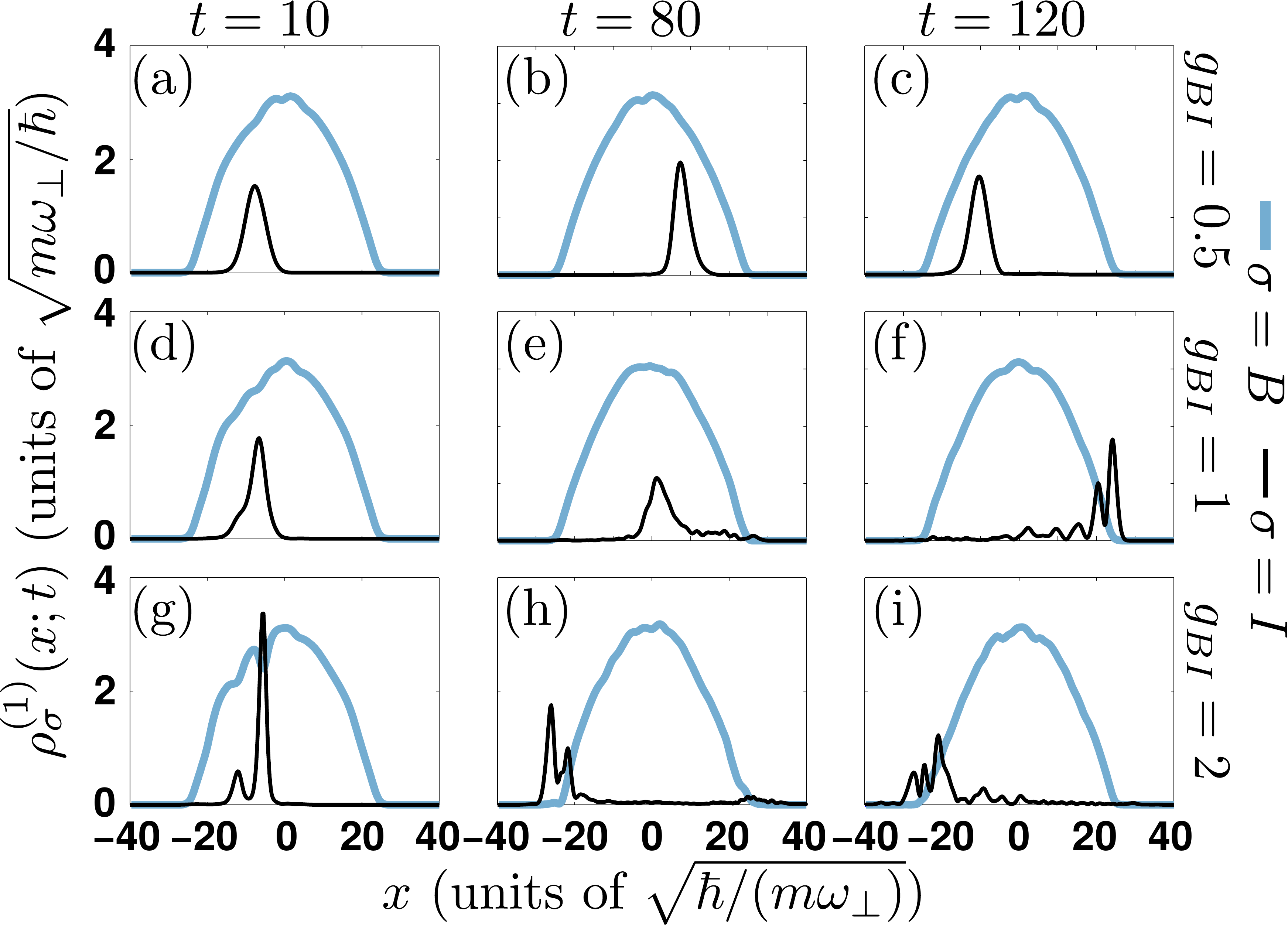}
     \caption{Snapshots of the $\sigma=B,I$ species one-body density at distinct time-instants (see legends) for a varying interspecies interaction strength 
     $g_{BI}$ (see legends). 
     The remaining system parameters are the same as in Fig. \ref{fig:2}.} 
     \label{fig:3}
\end{figure}

\begin{figure*}[ht]
 	\centering
  	\includegraphics[width=0.9\textwidth]{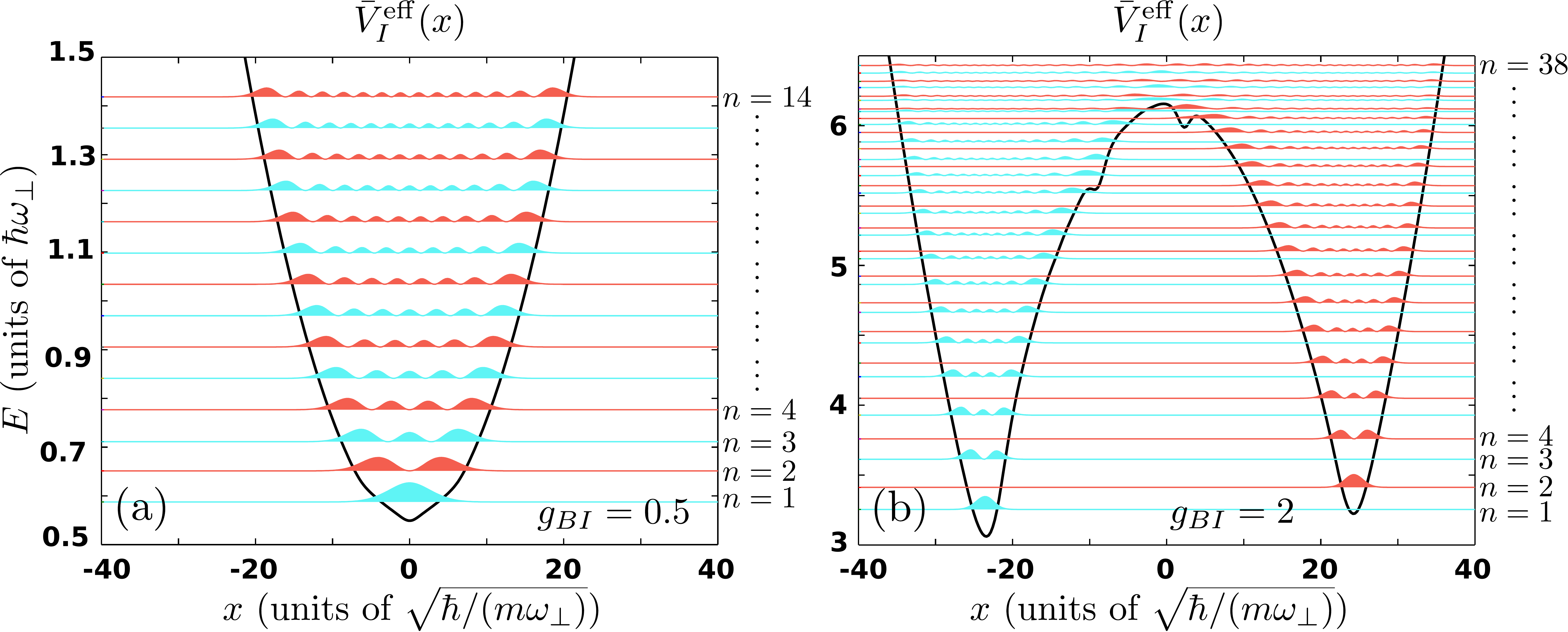}
     \caption{Time-averaged effective potential $\bar{V}_I^{eff}(x)$ [Eq. \ref{effective_potential_repulsive}] of the impurity for (a) weak and 
     (b) strong interspecies repulsions. 
     In all cases the densities of the impurity eigenenergies of $\bar{V}_I^{eff}(x)$ are depicted and the eigenstates are labeled with the principal quantum number $n$. 
     To obtain $\bar{V}_I^{eff}(x)$ we have used $T=150$. } 
     \label{fig:4}
\end{figure*}

Let us first inspect the dynamical response of the system upon considering an interspecies quench from $g_{BI}=0$ towards 
a finite positively valued $g_{BI}$.  
To achieve a spatially resolved description of the dynamics we resort to the time-evolution of the $\sigma$-species single-particle 
density $\rho^{(1)}_{\sigma}(x;t)$, see Fig. \ref{fig:2}. 
For a weak interspecies interaction quench, such that $g_{BI}<g_{BB}$, the impurity [see Fig. \ref{fig:2} (b)] performs almost perfect dipole 
oscillations of frequency $\omega_R\approx 0.07$ inside the bosonic medium. 
The deviation from perfect dipole oscillations, caused by the finite value of $g_{BI}$, is manifested in the shape of $\rho^{(1)}_{I}(x;t)$ since it becomes 
more wide when located close to the edges of $\rho^{(1)}_{B}(x;t)$ than the trap center, see Figs. \ref{fig:3} (a)-(c). 
The bosonic bath remains unperturbed to a large extent [Fig. \ref{fig:2} (a)] throughout the time-evolution, exhibiting small distortions at the core of its Thomas-Fermi 
cloud due to its interaction with the impurity. 
These distortions are directly evident in the instantaneous density profiles of $\rho^{(1)}_{B}(x;t)$ shown in Figs. \ref{fig:3} (a)-(c). 
It is also important to note here that the total external potential of the impurity can be well approximated by the time-averaged effective potential created by the 
external harmonic oscillator and the density of the bosonic bath. 
Such an effective potential picture affects the dynamics of the impurity in an essential manner only in the presence of an external 
trapping since in the homogeneous case the density of the bath is constant in space. 
More specifically, this effective potential \cite{Mistakidis_orth_cat,Hannes} reads 
\begin{equation}
 \bar{V}_I^{eff}(x)=\frac{1}{2}m\omega^2x^2+\frac{g_{BI}}{T}\int_0^T dt  \rho_B^{(1)}(x;t), \label{effective_potential_repulsive}
\end{equation} 
where $T$ denotes the considered evolution time. 
Here, we have used $T=150$. 
Let us also mention that this averaging process aims to eliminate the observed distortions on the instantanteous bosonic density $\rho^{(1)}_B(x;t)$, 
which is achieved in our case for $T>100$. 
These distortions are, of course, caused by the impurity motion and are mainly imprinted as sound waves, see e.g. Figs. \ref{fig:2} (c) and (e). 
This $\bar{V}_I^{eff}(x)$ at $g_{BI}=0.5$ corresponds to a modified harmonic oscillator potential and it is depicted in Fig. \ref{fig:4} (a) together with its first 
few single-particle eigenstates. 
In this case the impurity undergoes a dipole motion within $\bar{V}_I^{eff}(x)$ and predominantly resides in its energetically lowest-lying state, $n=1$. 

Performing a quench to stronger interspecies interaction strengths, e.g. $g_{BI}=1$, the collisional dynamics between the impurity 
and the bosonic bath [Figs. \ref{fig:2} (c), (d)] is drastically altered compared to the above-mentioned weakly interacting case [Figs. \ref{fig:2} (a)-(b)]. 
In particular, the prominent interspecies interactions greatly affect the motion of the impurity after the quench, see Fig. \ref{fig:2} (d). 
Inspecting $\rho_{I}^{(1)}(x;t)$ we observe that it undergoes an irregular oscillatory motion within the BEC. 
More specifically, it initially travels to the left edge of the bosonic bath where at $t\approx33$ it is reflected back towards the right edge possessing also 
a larger width compared to its initial one [see also Figs. \ref{fig:3} (d), (e)]. 
This change of the width of $\rho_{I}^{(1)}(x;t)$ is a direct effect of the interaction between the bosonic medium and the impurity and it becomes more pronounced 
when the impurity reaches the right edge of the bosonic bath and shows a multihump structure, see Fig. \ref{fig:2} (d) around $t\approx105$ and also Fig. \ref{fig:3} (f). 
This multihump structure of $\rho_{I}^{(1)}(x;t)$ suggests that the impurity populates a superposition of higher-lying excited states of the corresponding effective potential 
given by Eq. (\ref{effective_potential_repulsive}) as we shall discuss in more detail below. 
The bosonic medium becomes also perturbed due to its interaction with the impurity. 
As a result slight deviations from the initial Thomas-Fermi profile occur [Figs. \ref{fig:3} (d)-(f)] while $\rho_{B}^{(1)}(x;t)$ develops shallow density dips at the 
instantaneous location of the density hump of $\rho_{I}^{(1)}(x;t)$, thus imprinting the motion of the impurity. 

Entering to stronger postquench interspecies interactions, e.g. $g_{BI}=2$, which satisfy $g_{BI}>g_{BB}$ reveals a completely different dynamical response of both 
species, see Figs. \ref{fig:2} (e), (f). 
Remarkably enough the impurity travels towards the left edge of the bosonic bath where it remains locked while exhibiting an oscillatory behavior of negligible amplitude for $t>110$. 
The fact that the impurity escapes from the bosonic gas and undergoes damped oscillations around its Thomas-Fermi radius is 
reminiscent of an  orthogonality catastrophe phenomenon \cite{Mistakidis_orth_cat}.
Indeed, it has been showcased that upon considering an interspecies interaction quench of a zero velocity 
impurity atom immersed in a bosonic gas to strong repulsions, the structure factor \cite{Cetina_interferometry} of the quasiparticle becomes zero and simultaneously the density of the impurity 
resides at the edges of the BEC \cite{Mistakidis_orth_cat}. 
However, an important difference between a zero velocity and a moving impurity is that in the former case after the quench $\rho_{I}^{(1)}(x;t)$ breaks into two fragments, a behavior 
that does not occur herein. 
Note also that due to its interaction with the BEC background the impurity emits some small portions of density when it is located well inside $\rho_{B}^{(1)}(x;t)$, see for 
instance the dashed white rectangles in Fig. \ref{fig:2} (f) and the small amplitude density hump of $\rho_{I}^{(1)}(x;t)$ in Fig. \ref{fig:3} (g). 
When the impurity reaches the left edge of the Thomas-Fermi cloud its $\rho_{I}^{(1)}(x;t)$ develops a multihump structure which indicates that it resides in a superposition of 
several energetically higher-lying excited states of the effective external potential. 
In this case of strong $g_{BI}$ the effective potential introduced in Eq. (\ref{effective_potential_repulsive}) is an asymmetric double-well potential which 
exhibits an energy offset between the left and right wells. 
We remark here that the effective potential approximation provides for these strong interactions only a very approximate but rather intuitive picture of the impurity dynamics. 
However, $\rho_{I}^{(1)}(x;t)$ shown in Fig. \ref{fig:2} (f) and Figs. \ref{fig:3} (g)-(i) has been obtained within the full many-body approach described in Section \ref{ML_ansatz}. 
The resulting $\bar{V}_{I}^{eff}(x)$ and its first few single-particle eigenstates are illustrated in Fig. \ref{fig:4} (b). 
In terms of this effective picture for $t>100$, the impurity is trapped in the left well of $\bar{V}_{I}^{eff}(x)$ where it predominantly occupies a superposition of 
the $n=1$, $n=3$ and $n=5$ eigenstates. 
Furthermore, the motion of the impurity leaves its fingerprints also in the BEC background which as a result becomes perturbed. 
Indeed, the Thomas-Fermi cloud is disturbed as it can be seen from the corresponding instantaneous density profiles presented in Figs. \ref{fig:3} (g)-(i) and in particular when 
$\rho_{I}^{(1)}(x;t)$ is well inside $\rho_{B}^{(1)}(x;t)$ [e.g. see Fig. \ref{fig:3} (g)] the latter develops density dips at the instantaneous location of $\rho_{I}^{(1)}(x;t)$. 
Note also that the small distortions appearing in the spatial region of the barrier of the effective double-well potential depicted in Fig. \ref{fig:4} (b) are caused by the existence 
of beyond mean-field corrections at the core of $\rho_{B}^{(1)}(x;t)$.

\subsection{Mean Position of the Impurity}\label{sec:position_repulsive}

\begin{figure*}[ht]
 	\centering
  	\includegraphics[width=0.8\textwidth]{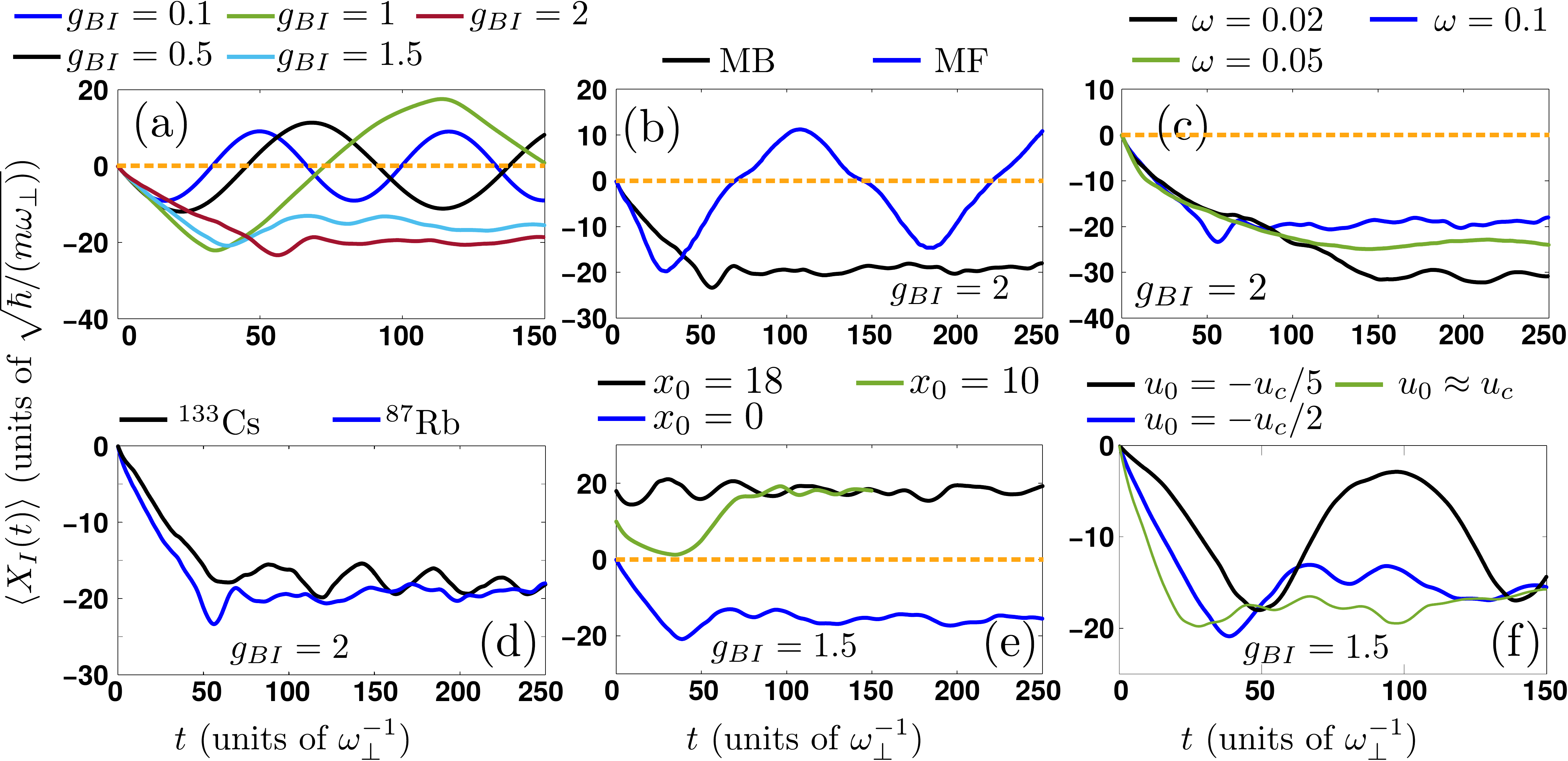}
     \caption{(a) Time-evolution of the position of the impurity, $\braket{X_I(t)}$, for different interspecies interaction strengths (see legend).  
     (b) $\braket{X_I(t)}$ at $g_{BI}=2$ within the mean-field (MF) and the many-body (MB) approach. 
     Dynamics of $\braket{X_I(t)}$ for distinct (c) trapping frequencies $\omega$ of the bosonic gas and (d) masses $m_I$, (e) initial positions $x_0$ and (f) 
     initial velocities $u_0$ of the impurity. 
     The straight dashed lines in (a), (b), (c) and (e) provide a guide to the eye for $\braket{X_I(t)}=0$. 
     In each case all other system parameters are kept fixed and are the same as in Fig. \ref{fig:2}.} 
     \label{fig:5}
\end{figure*}

To examine the dependence of the dynamical response of the impurity on the distinct system parameters we next monitor its motion by calculating 
its mean position $\braket{X_I(t)}$ [see also Eq. (\ref{mean_position})] during the dynamics. 
We first investigate the motion of a subsonic impurity with initial velocity $u_0=-u_c/2$ and the quench is performed when it is located at $x_0=0$. 
Figure \ref{fig:5} (a) shows $\braket{X_I(t)}$ for different postquench interspecies interaction strengths. 
In line with our discussion in Section \ref{sec:density_repulsive}, we observe that for $g_{BI}<g_{BB}$ the impurity oscillates within the BEC background but with an increasing 
period for a larger $g_{BI}$, e.g. compare $\braket{X_I(t)}$ between $g_{BI}=0.1$ and $g_{BI}=0.5$. 
However, for quench amplitudes characterized by $g_{BI}>g_{BB}$ the impurity moves towards the left edge of the bosonic bath and subsequently equilibrates [Fig. \ref{fig:2} (f)], 
e.g. see $\braket{X_I(t)}$ at $g_{BI}=1.5$ for $t>50$. 

A natural question that arises is whether this latter behavior of the impurity, namely equilibration at the edge of the BEC background, is an effect of the 
inclusion of the correlations into the dynamics. 
To address this question we next 
present $\braket{X_I(t)}$ at $g_{BI}=2$ in Fig. \ref{fig:5} (b) within the fully many-body approach [Eq. (\ref{Eq:WF})] and the mean-field approximation [Eq. (\ref{Eq:MF})]. 
Evidently, $\braket{X_I(t)}$ within the mean-field approximation exhibits an oscillatory behavior for long evolution times meaning that the impurity remains independently of $g_{BI}$ 
well inside the BEC background. 
This sharp contrast of the behavior of $\braket{X_I(t)}$ between the many-body and the mean-field treatments occurring at large $g_{BI}$ evinces that the observed equilibration of 
the impurity at the edge of the bosonic bath is a direct effect of the presence of correlations. 
Moreover, this behavior of $\braket{X_I(t)}$ taking place at strong $g_{BI}$ occurs even for a decreasing trapping frequency e.g. $\omega=0.05$ 
(at larger evolution times) as shown in Fig. \ref{fig:5} (c). 
The fact that the phenomenon occurs for larger evolution times can be attributed to the fact that for a decreasing $\omega$, and thus tending to the untrapped case, 
the corresponding Thomas-Fermi radius of the BEC becomes larger and therefore the impurity needs to travel a longer distance until it reaches the edge of the BEC cloud. 
This behavior is a direct manifestation of the effect of the strength of the external trapping on the equilibration time of the moving impurity at strong 
interspecies repulsions. 
Let us also remark in passing that at short evolution times ($0<t<40$ in Fig. \ref{fig:5} (c)) the trajectory, and thus also the corresponding velocity, of the impurity is 
independent of the harmonic oscillator frequency.  
This is an expected result since for these short evolution times the impurity lies well inside the bosonic gas and thus experiences an almost 
homogeneous environment. 

Interestingly, the equilibration of the impurity occurring at strong repulsions persists also for a heavier impurity atom as depicted in Fig. \ref{fig:5} (d). 
Here we consider a $^{87}$Rb bosonic gas and a single $^{133}$Cs impurity at the hyperfine states $\Ket{F=1, m_F=0}$ and $\Ket{F=3, m_F=2}$ respectively both trapped 
in an external harmonic oscillator of the same frequency \cite{Hohmann_Rb_Cs,Spethmann_Rb_Cs}. 
Also, the initial momentum of the subsonic impurity is kept fixed in both cases. 
As it can be seen, $\braket{X_I(t)}$ reaches the edge of the BEC at almost the same time scales in both mixtures but the $^{133}$Cs atom remains inside the Thomas-Fermi 
radius to a larger extent than the $^{87}$Rb one. 
This is an expected result since the velocity $u_0$ of the $^{133}$Cs impurity is smaller than the corresponding $^{87}$Rb one because $m_{Cs}>m_{Rb}$.    

Focusing on the strongly interacting regime, e.g. $g_{BI}=2$, we next inspect $\braket{X_I(t)}$ by considering the interaction quench at different locations $x_0$ of the impuritys' 
motion with respect to the trap center, see Fig. \ref{fig:5} (e). 
As it can be deduced, $\braket{X_I(t)}$ exhibits a saturated behavior independently of $x_0$. 
Notice also here that for $x_0>0$ the impurity is repelled by the bosonic cloud to the opposite direction of its motion and finally reaches 
the right edge of the BEC background, e.g. see $\braket{X_I(t)}$ for $x_0=10$. 
This is, of course, a manifestation of the exerted force by the BEC on the impurity. 
Accordingly, we can infer that the density of the impurity approaches selectively the smaller distant edge of the bosonic bath in terms of its prequench position. 
Note that this result is in contrast to the behavior of a zero velocity impurity whose density at such strong repulsions breaks into two fragments which exhibit a dissipative oscillatory 
motion around the edges of the bosonic gas \cite{Mistakidis_orth_cat}. 

Next, we examine the dependence of the motion of the impurity in the strongly interacting regime, $g_{BI}=2$, on its initial velocity $u_0$ when the quench is performed 
at position $x_0=0$. 
Figure \ref{fig:5} (f) illustrates $\braket{X_I(t)}$ for initial velocities $u_0\approx-(1/5)u_c$ (subsonic), $-(1/2)u_c$ (subsonic) and $-u_c$ (sonic) with $u_c\approx1.74$ 
being the speed of sound of the BEC background. 
We deduce that for an increasing initial velocity, such that $u_0\to -u_c$, the impurity reaches faster the left edge of the bosonic bath where it subsequently equilibrates. 
Note that this behavior of $\braket{X_I(t)}$ for $u_0\approx -u_c$ is in contrast to the long-lived oscillations reported in homogeneous settings \cite{Flutter,Flutter1} but 
for supersonic ($u_0\gg u_c$) impurities. 
However for $u_0\ll u_c$, e.g. $u_0=(1/5)u_c$, the impurity performs oscillations through the BEC of a much slower decaying amplitude when compared to the previous cases. 
This behavior is caused due to its small velocity which generates a lesser amount of excitations to the BEC as compared to large $u_0$.

\subsection{Degree of Entanglement}\label{sec:entanglement_repulsive} 

To quantify the correlated nature of the collisional dynamics between the impurity and the BEC we next measure the degree of entanglement or interspecies correlations by 
employing the Von-Neumann entropy $S_{VN}(t)$ [Eq. (\ref{eq:entropy})]. 
Recall that $S_{VN}(t)\neq0$ signifies the presence of interspecies entanglement, otherwise the system is non-entangled \cite{mistakidis_phase_sep}. 

The dynamics of $S_{VN}(t)$ is shown in Fig. \ref{fig:6} (a) following an interspecies interaction quench for different values of $g_{BI}$. 
As it can be seen $S_{VN}(t=0)=0$ since for the initial state of the system $g_{BI}=0$. 
However, after the quench $S_{VN}(t)$ acquires finite values thus indicating the presence of interspecies correlations. 
For weak postquench interactions, e.g. $g_{BI}=0.3$, there is only a small amount of interspecies correlations since $S_{VN}(t)$ is 
suppressed taking very small values. 
Recall that in this case the impurity performs dipole oscillations within the BEC, see also Fig. \ref{fig:2} (b). 
On the contrary, for stronger postquench interactions such as $g_{BI}=1.5$ $S_{VN}(t)$ increases rapidly at the initial stages of the dynamics where the 
impurity resides within the BEC while for later times at which the impurity equilibrates at the edge of the bosonic gas $S_{VN}(t)$ tends to saturate to a certain finite value. 
This behavior of $S_{VN}(t)$ indicates that the underlying many-body state [Eq. (\ref{Eq:WF})] is strongly entangled. 
It is worth mentioning that for strong repulsions where the impurity essentially escapes from the BEC, e.g. at $g_{BI}=2$ for $t>80$ [Fig. \ref{fig:2} (f)], 
suggesting a break down of the quasiparticle picture $S_{VN}(t)$ acquires an almost constant value [Fig. \ref{fig:6} (a)]. 
Also stronger postquench interspecies interactions, $g_{BI}$, result in larger values of $S_{VN}(t)$.

\subsection{Interspecies Energy Transfer}\label{sec:energy_repulsive} 

To further understand the nonequilibrium dynamics of the impurity immersed in the BEC background for a different postquench $g_{BI}$, 
below we analyze the distinct energy contributions of the bosonic mixture \cite{Mistakidis_orth_cat,Mistakidis_two_imp_ferm,Nielsen}. 
The normalized energy of the BEC corresponds to $E_B(t)=\braket{\Psi(t)|\hat{T}_B+\hat{V}_B(x)+\hat{H}_{BB}|\Psi(t)}-\braket{\Psi(0)|\hat{T}_B+\hat{V}_B(x)+\hat{H}_{BB}|\Psi(0)}$, and 
for the impurity is $E_I(t)=\braket{\Psi(t)|\hat{T}_I+\hat{V}_I(x)|\Psi(t)}$. 
Moreover, the interspecies interaction energy is $E_{BI}(t)=\braket{\Psi(t)|\hat{H}_{BI}|\Psi(t)}$. 
In this notation, $\hat{T}_{\sigma}=-\int dx \hat{\Psi}^{\sigma \dagger}(x)\frac{\hbar^2}{2 m} (\frac{d}{dx^{\sigma}})^2 \hat{\Psi}^{\sigma}(x)$ and 
$\hat{V}_{\sigma}(x)=\int dx \hat{\Psi}^{\sigma \dagger}(x)\frac{1}{2} m \omega^2 x_{\sigma}^2 \hat{\Psi}^{\sigma}(x)$ denote the kinetic and the potential energy operators 
of the $\sigma=B,I$ species respectively. 
Also, $\hat{H}_{BB}=g_{BB} \int dx~\hat{\Psi}^{B \dagger}(x) \hat{\Psi}^{B \dagger}(x) \hat{\Psi}^{B} (x)\hat{\Psi}^{B}
(x)$ and $\hat{H}_{BI}=g_{BI}\int dx~\hat{\Psi}^{B \dagger}(x) \hat{\Psi}^{I \dagger}(x) \hat{\Psi}^{I}(x)\hat{\Psi}^{B}(x)$ refer to the operators of the intra- and 
interspecies interactions with $\hat{\Psi}^{\sigma} (x)$ being the $\sigma$-species field operator. 
\begin{figure*}[ht]
 	\centering
  	\includegraphics[width=1.0\textwidth]{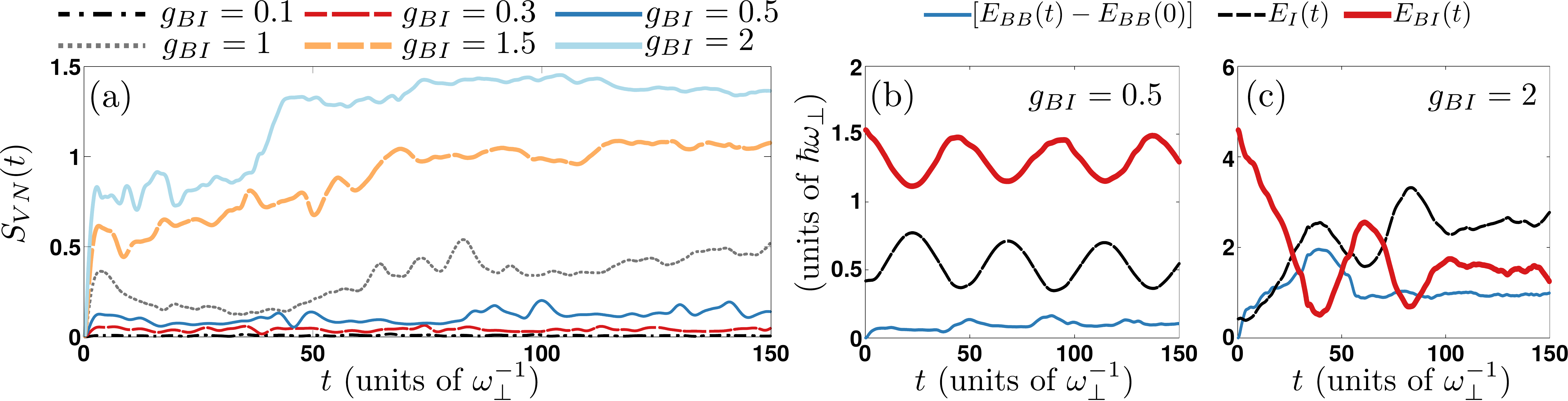}
     \caption{(a) Evolution of the Von-Neumann entropy between the two species for varying interspecies interaction strength (see legend).  
     (b) Expectation value of the energy of the bosonic bath $E_{B}(t)$, the impurity $E_{I}(t)$ and their interspecies interaction energy $E_{BI}(t)$ following an 
     interspecies interaction quench from $g_{BI}=0$ to (b) $g_{BI}=0.5$ and (c) $g_{BI}=2$. 
     Note the different energy scales of (b) versus (c). 
     The remaining system parameters are the same as in Fig. \ref{fig:2}.} 
     \label{fig:6}
\end{figure*} 

The dynamics of each of the above-described energy contributions is presented in Figs. \ref{fig:6} (b), (c) upon considering a quench towards weak and strong 
interspecies repulsive interactions respectively. 
Focusing on weak postquench interactions [Fig. \ref{fig:6} (b)], e.g. $g_{BI}=0.5$, the energy of the impurity $E_I(t)$ and the interspecies interaction energy $E_{BI}(t)$ 
exhibit an oscillatory behavior. 
The energy of the bath, $E_{B}(t)$, slightly increases and $E_{B}(t)<E_{I}(t)<E_{BI}(t)$ holds in the course of the evolution. 
In particular, $E_I(t)$ is minimized at the time-intervals where the impurity is close to the trap center and it is maximized when the impurity travels towards the edges 
of the BEC. 
Accordingly, $E_{BI}(t)$ oscillates out-of-phase with $E_I(t)$ since the interspecies interaction is stronger when the impurity is close to the trap center (where $E_I$ is small) and vice versa. 
Moreover, the fact that $E_{B}(t)$ increases to a minor extent during the dynamics suggests that the impurity devolves a small amount of energy to the BEC. 
This process is captured by the very weakly decaying amplitude of oscillations of $E_I(t)$. 

Referring to strong interspecies interactions, e.g. $g_{BI}=2$ shown in Fig. \ref{fig:6} (c) the dynamical behavior of all energy contributions is drastically altered 
when compared to their weakly interacting counterparts [compare Figs. \ref{fig:6} (b) and (c)]. 
At $0<t<40$, $E_{BI}(t)$ reduces while $E_{B}(t)$ and $E_{I}$ increase. 
Indeed, within this time interval the impurity density moves to the left edge of the BEC [Fig. \ref{fig:2} (b)] with a large kinetic energy and as a result dissipates 
energy to the latter. 
For $40<t<100$, $E_{I}(t)$ and $E_{BI}(t)$ oscillate out-of-phase with respect to one another and in particular $E_{I}(t)$ overall increases while performing small 
amplitude oscillations. 
Note that in this time interval $\rho^{(1)}_{I}(x;t)$ oscillates around the left boundary of $\rho^{(1)}_{B}(x;t)$ and still weakly interacts 
with the BEC. 
Simultaneously, $E_{B}(t)$ increases when the impurity resides to a large extent in the bosonic bath and decreases when $\rho^{(1)}_{I}(x;t)$ is located 
at the left edge. 
As a result the impurity transfers a part of its energy to the bosonic gas. 
Similar energy exchange processes between the impurity and the host atoms have already been observed e.g. in Refs. \cite{Mistakidis_orth_cat,Nielsen}. 
Deeper in the evolution, $t>100$, all energy components acquire an almost constant value with $E_{B}(t)< E_{BI}(t)<E_{I}(t)$. 
Recall that for $t>100$ $\rho^{(1)}_{I}(x;t)$ resides at the left edge of $\rho^{(1)}_{B}(x;t)$ and therefore the interaction between the two species is drastically reduced.

\section{Quench Dynamics Towards Attractive Interactions}\label{sec:quench_attractive} 

Next, we explore the out-of-equilibrium dynamics of a subsonically moving impurity ($N_I=1$) immersed within a harmonically trapped BEC ($N_B=100$) 
upon considering an interspecies interaction quench towards attractive interactions. 
As in the previous Section \ref{sec:quench_repulsive}, the BEC is initially prepared into its ground state with $g_{BB}=1$ having a Thomas-Fermi profile 
of radius $R_{TF}\approx25$. 
The subsonic impurity is initially modeled as a coherent state [Eq. (\ref{coherent_state})] with a velocity $u_0=-u_c/2$ and the impurity-BEC interaction is zero ($g_{BI}=0$) at $t=0$. 
To induce the dynamics we perform at $t=0$ a quench to negative $g_{BI}$ interaction strengths when the impurity is at $x_0=0$.

\subsection{Density Evolution and Effective Picture}\label{sec:density_attractive}

\begin{figure*}[ht]
 	\centering
  	\includegraphics[width=0.9\textwidth]{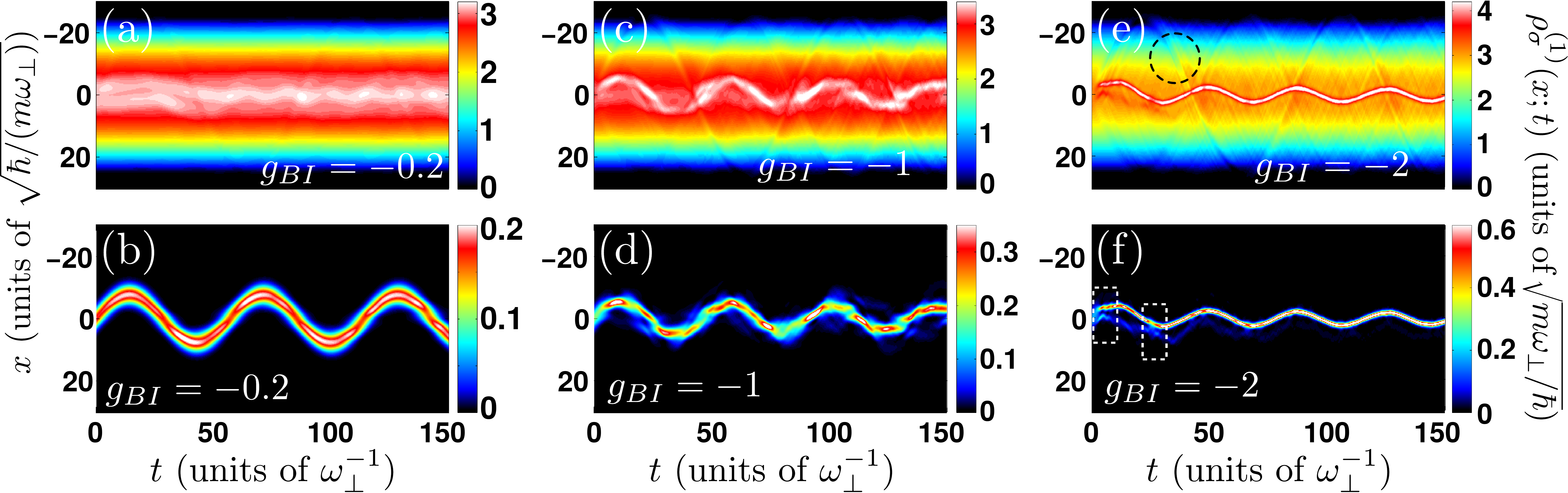}
     \caption{One-body density evolution of the bath (upper panels) and the impurity (lower panels) for different attractive interspecies interaction strengths 
     $g_{BI}$ (see legends). 
     The system consists of $N_{B}=100$ bosons initialized in their ground state with $g_{BB}=1$ and $N_I=1$ impurity atoms being in a coherent state 
     which is located at $x_0=0$ and possesses an initial velocity $u_0=-u_c/2$. 
     The dashed circle in (e) indicates the emission of sound waves in the bosonic gas, while the dashed rectangles in (f) mark the emission 
     and re-collision of a small density portion of the impurity. 
     Both species are trapped in an external harmonic oscillator of frequency $\omega=0.1$. 
     To induce the dynamics at $t=0$ we quench the interspecies coupling constant from $g_{BI}=0$ to a finite negative value (see legends).} 
     \label{fig:7}
\end{figure*}

\begin{figure}[ht]
 	\centering
  	\includegraphics[width=0.45\textwidth]{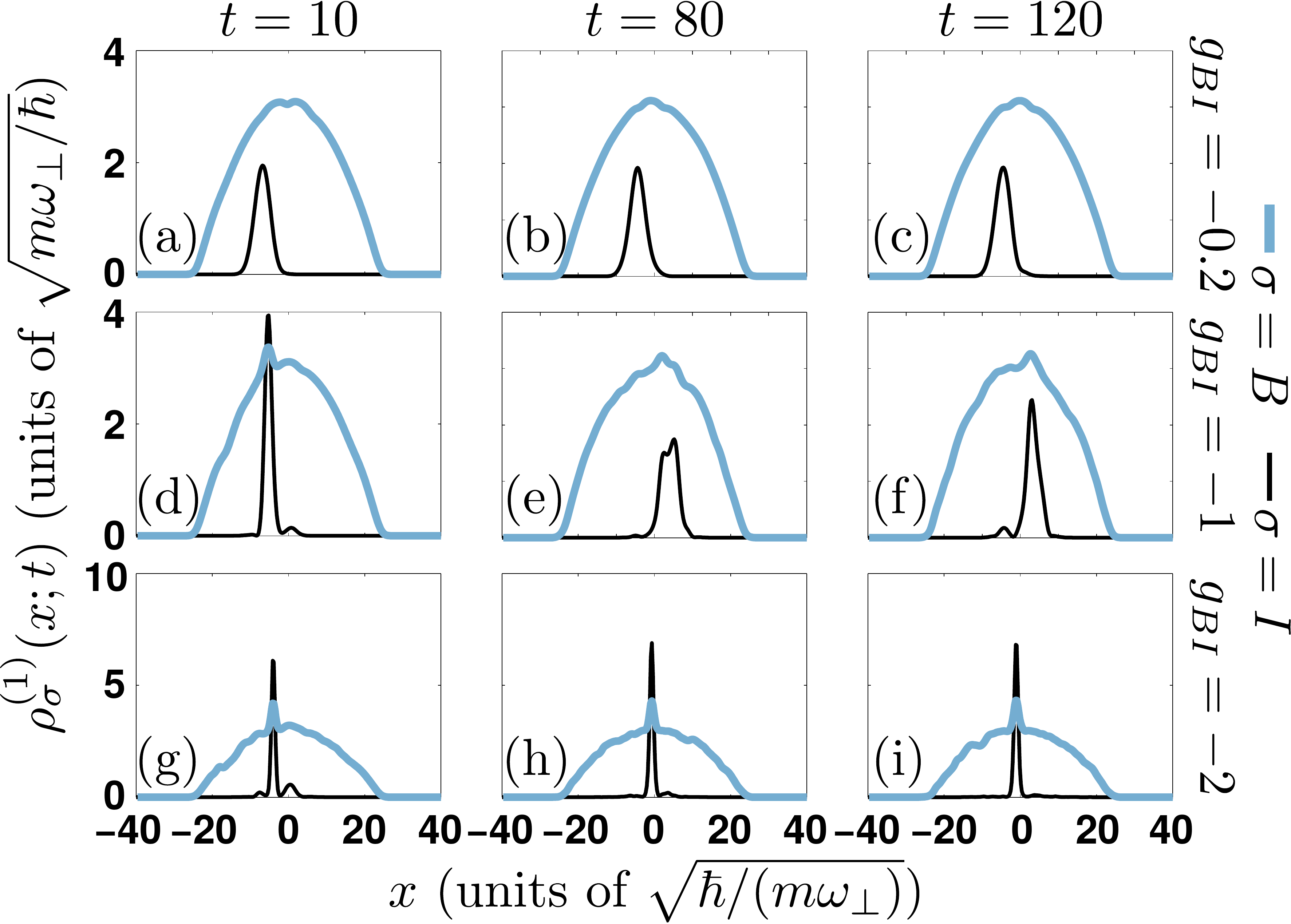}
     \caption{Density profiles of the $\sigma=B,I$ species at various time-instants (see legends) of the evolution for distinct interspecies interaction strengths 
     $g_{BI}$ (see legends). 
     The remaining system parameters are the same as in Fig. \ref{fig:8}.} 
     \label{fig:8}
\end{figure}

To unveil the dynamical response of the system after an interspecies interaction quench to attractive coupling strengths we resort to the time-evolution 
of the $\sigma$-species single-particle density $\rho_{\sigma}^{(1)}(x;t)$. 
The emergent evolution of $\rho_{\sigma}^{(1)}(x;t)$ is illustrated in Fig. \ref{fig:7} for distinct postquench interspecies interactions ranging from weak to strong 
negative values. 
Focusing on weak postquench negative interactions, $g_{BI}=-0.2$ [Figs. \ref{fig:7} (a), (b)], we observe that due to the small $g_{BI}$ $\rho_{I}^{(1)}(x;t)$ oscillates 
within $\rho_{B}^{(1)}(x;t)$ with an almost fixed amplitude, see also Figs. \ref{fig:8} (a)-(c), and frequency $\omega_{osc}\approx 0.11$. 
As a result of the motion of the impurity and the weak $g_{BI}$ the Thomas-Fermi cloud is slightly distorted and in particular faint density humps built upon 
$\rho_{B}^{(1)}(x;t)$ at the location of $\rho_{I}^{(1)}(x;t)$ [hardly visible in Fig. \ref{fig:7} (a)].  

\begin{figure}[ht]
 	\centering
  	\includegraphics[width=0.45\textwidth]{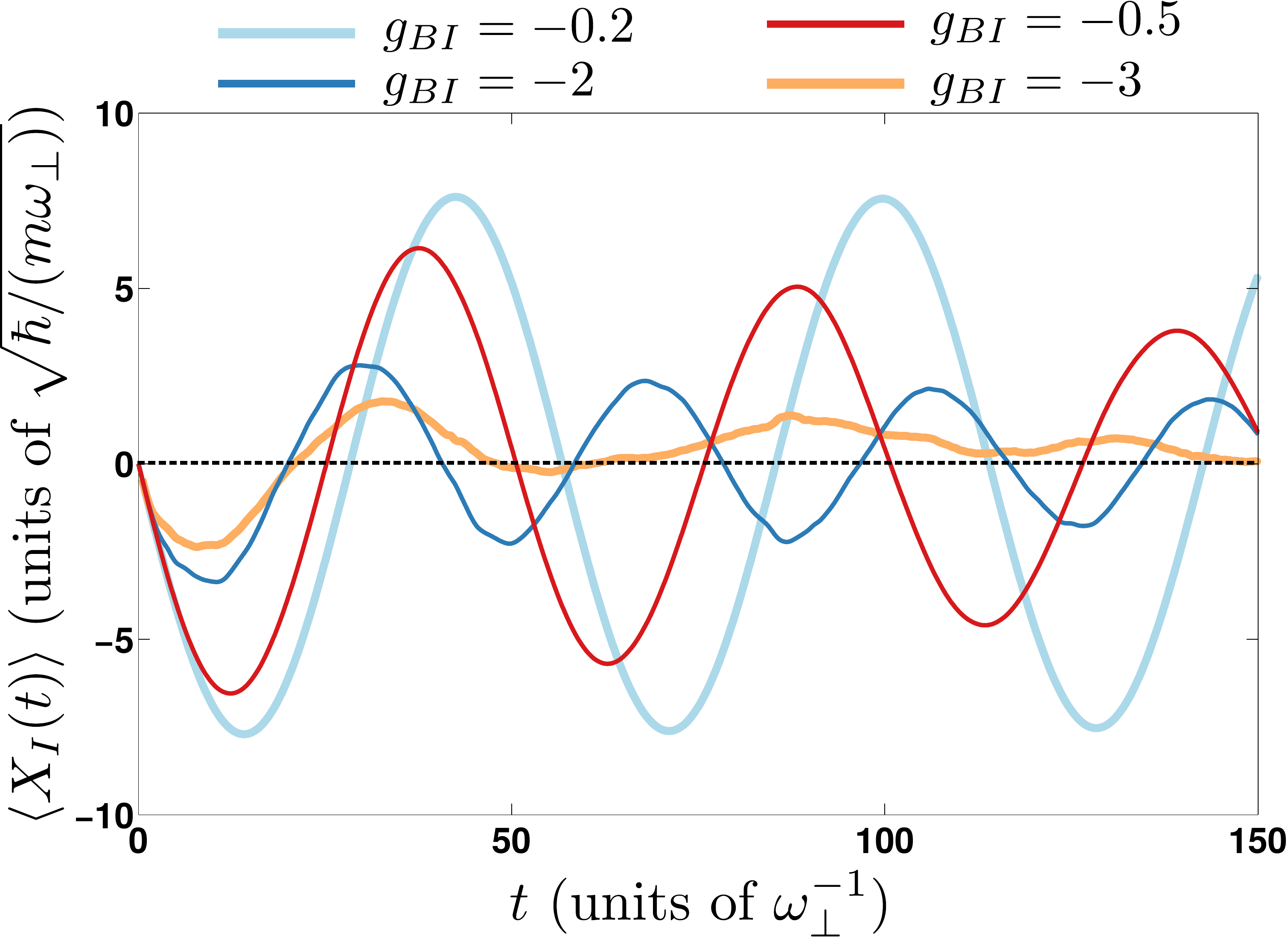}
     \caption{Position of the impurity in the course of the dynamics for different attractive interspecies interaction 
     strengths (see legend). 
     The remaining system parameters are kept fixed and are the same as in Fig. \ref{fig:8}.} 
     \label{fig:9}
\end{figure}

For stronger negative interspecies interactions, e.g. $g_{BI}=-1$, $\rho_{I}^{(1)}(x;t)$ undergoes an oscillatory motion of decaying amplitude within the BEC 
background and a larger frequency $\omega_{osc}\approx0.14$ compared to the $g_{BI}=-0.2$, see Fig. \ref{fig:7} (d). 
Due to the finite $g_{BI}$ the motion of $\rho_I^{(1)}(x;t)$ in turn results in the development of a density hump on $\rho_B^{(1)}(x;t)$ at the instantaneous 
position of the impurity, see Fig. \ref{fig:7} (c) and Figs. \ref{fig:8} (d)-(f). 
This decaying amplitude oscillatory behavior of the impurity persists and becomes more evident for stronger attractive $g_{BI}$, 
compare Figs. \ref{fig:7} (d) and (f). 
Note also the additional modulations of the density peak of the impurity caused by its collisions with the excitations of the bosonic gas. 
The above-mentioned behavior of $\rho_{I}^{(1)}(x;t)$ can be directly captured by inspecting the dynamics of the mean position of the impurity 
$\braket{X_I(t)}$ for varying $g_{BI}$ shown in Fig. \ref{fig:9}. 
Indeed, we can deduce that $\braket{X_I(t)}$ oscillates with a decaying amplitude in time which is more pronounced deeper in the attractive regime of 
interactions, compare $\braket{X_I(t)}$ for $g_{BI}=-0.5$ and $g_{BI}=-2$. 
This attenuation of the oscillation amplitude of $\braket{X_I(t)}$ is a direct effect of the presence of interspecies interactions and the underlying energy transfer 
process from the impurity to the bath, see also the discussion below and Refs. \cite{Mistakidis_eff_mass,Nielsen}. 
Also, $\rho_{I}^{(1)}(x;t)$ having a sech-like shape tends to be more localized for larger negative values of $g_{BI}$ [see Figs. \ref{fig:8} (g)-(i)], 
a result that holds equally for the corresponding density hump building upon $\rho_{B}^{(1)}(x;t)$ [Fig. \ref{fig:7} (e)]. 
The latter density hump being directly discernible in $\rho_{B}^{(1)}(x;t)$ is essentially an imprint of the impurity motion inside the BEC. 
Another interesting observation here is that at large $\abs{g_{BI}}$ the system is strongly correlated and the BEC background is highly excited, as can be 
inferred from the emission of a large amount of sound waves, see for instance the dashed black circle in Fig. \ref{fig:7} (e) and the discussion below. 
Such a sound wave emission has been extensively reported during the motion of a Gaussian barrier inside a BEC within [e.g. see Refs. \cite{sound_wave,sound_wave1}] and 
beyond \cite{Katsimiga_diss_flow} the mean-field approximation. 
In the present investigation the impurity plays, of course, the role of the Gaussian barrier. 
According to these studies the motion of the impurity locally perturbs the initial zero phase of the BEC leading to the formation of small amplitude phase disturbances 
that lead to sound waves. 
A similar mechanism takes place also herein where we can identify the existence of sound waves by measuring their velocity at the center of the trap 
being larger than $0.95 u_c$. 

\begin{figure*}[ht]
 	\centering
  	\includegraphics[width=0.9\textwidth]{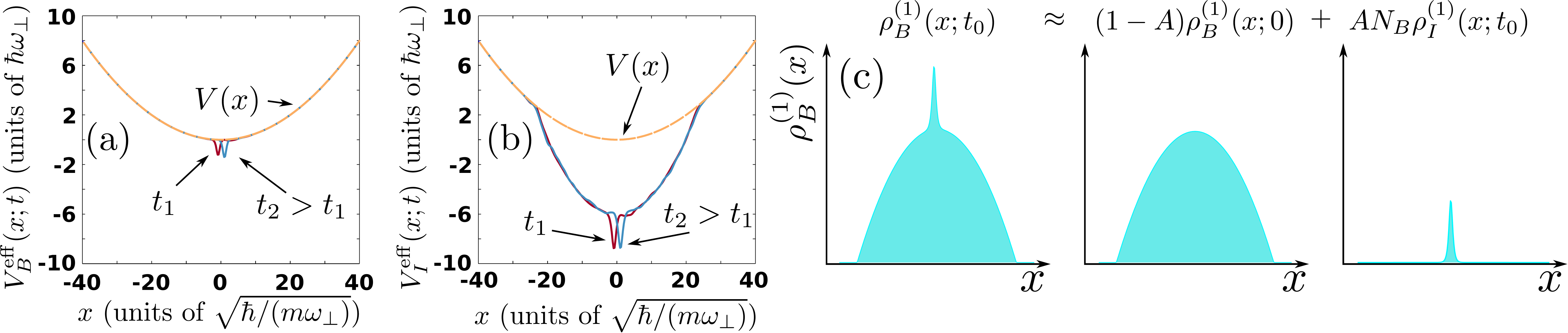}
     \caption{Effective potential experienced by (a) the BEC background [Eq. \ref{effective_pot_bath}] and (b) the impurity particle [Eq. \ref{effective_pot_impurity}] for $g_{BI}=-2$ 
     at different time-instants $t_1$, $t_2$ of the dynamics. 
     (c) The single-particle density of the BEC background (left panel) at a certain time-instant $t_0$ decomposed into its corresponding Thomas-Fermi profile (central panel) 
     and a $\sech$-shaped wavepacket (right panel). 
     $A$ is a real-valued parameter accounting for the deformation of $\rho^{(1)}_{B}(x;t_0)$ from the Thomas-Fermi profile.} 
     \label{fig:10}
\end{figure*} 

The above-described dynamical response of the impurity and the BEC taking place at these negative interspecies interactions can be qualitatively understood by 
invoking an effective potential picture \cite{Mistakidis_orth_cat,Hannes}. 
Indeed, the effective potential acting on the BEC consists of the external harmonic oscillator and the instantaneous density of the impurity namely  
\begin{equation}
 V_B^{eff}(x,t)= V(x)-\abs{g_{BI}} \rho_I^{(1)}(x;t). \label{effective_pot_bath}  
\end{equation} 
A schematic illustration of $V_B^{eff}(x,t)$ at $g_{BI}=-2$ is shown in Fig. \ref{fig:10} (a) at two distinct time-instants of the evolution. 
We deduce that $V_B^{eff}(x,t)$ corresponds to an harmonic oscillator like potential possessing also a dip, at the momentary position of the impurity, 
which is characterized by negative energies. 
This latter attractive part of $ V_B^{eff}(x,t)$ is responsible for the observed density hump appearing in the dynamics of $\rho_B^{(1)}(x;t)$. 
Accordingly, the effective potential of the impurity is created by the external harmonic oscillator $V(x)$ and the single-particle density of the BEC. 
We remark that since $\rho_B^{(1)}(x;t)$ is greatly affected by the impurity motion, a time-average effective potential cannot adequately capture 
the dynamics of the impurity. 
In particular, the effective potential of the impurity reads 
\begin{equation}
 V_I^{eff}(x,t)=V(x)-\abs{g_{BI}} \rho_B^{(1)}(x;t). \label{effective_pot_impurity} 
\end{equation} 
Figure \ref{fig:10} (b) presents $V_I^{eff}(x,t)$ calculated at $g_{BI}=-2$ for two different times in the course of the dynamics. 
As shown, $V_I^{eff}(x,t)$ is a deformed attractive harmonic oscillator potential having an additional dip around $x\approx0$ due to the presence of 
the density hump building upon $\rho_B^{(1)}(x;t)$. 
This attractiveness of $V_I^{eff}(x,t)$ causes the localized sech-like shape of $\rho_I^{(1)}(x;t)$ located around the aforementioned 
additional potential dip. 
Most importantly, the observed distinct features of the impurity occurring for stronger attractive interactions can be explained via the behavior 
of the constructed $V_I^{eff}(x,t)$. 
Indeed, for increasing $\abs{g_{BI}}$ the effective frequency of $V_I^{eff}(x,t)$ is larger and $V_I^{eff}(x,t)$ becomes more attractive. 
The former property of $V_I^{eff}(x,t)$ accounts for the decreasing oscillation period of $\braket{X_I(t)}$ for larger $\abs{g_{BI}}$. 
Additionally, the increasing attractiveness of $V_I^{eff}(x,t)$ is responsible for the reduced width of $\rho_I^{(1)}(x;t)$ for a larger $\abs{g_{BI}}$ and 
thus its increasing localization tendency, e.g. compare Figs. \ref{fig:7} (d) and (f). 

To showcase the interconnection between $V^{eff}_I(x;t)$ and $V^{eff}_B(x;t)$ we approximatively decompose the one-body density
of the BEC at time $t_0$ according to
\begin{equation}
\rho^{(1)}_B(x;t_0) \approx (1-A) \rho^{(1)}_B(x;0) + A N_B \rho^{(1)}_I(x;t_0).
\label{eff_expansion}
\end{equation} 
The first term in Eq. (\ref{eff_expansion}) corresponds to the unperturbed BEC in the absence of the impurity.
The second term provides a correction to $\rho^{(1)}_B(x;t_0)$ stemming from the interspecies interaction according to $V_B^{eff}(x,t)$ 
[see Eq. (\ref{effective_pot_bath}) and also Fig. \ref{fig:10} (c)]. 
Also, $A$ is a real valued parameter bounded in the interval $[0,1]$. 
In the sense of Eq. (\ref{eff_expansion}) positive values of $A$ encode the back-action of $\rho^{(1)}_I(x;t_0)$ on the density of the BEC. 
The latter, in turn, forms the effective potential $V^{eff}_I(x;t)$ as dictated by Eq. (\ref{effective_pot_impurity}) which accordingly 
determines $\rho^{(1)}_I(x;t_0)$. 
Indeed, Fig. \ref{fig:7} (c)-(d) and \ref{fig:8} (g)-(h) indicate that this correction proportional to $A$ is sizable especially in the case of strong
attractive interactions, e.g. $g_{BI}=-2$. 
This is in sharp contrast to an impurity repulsively interacting with a BEC where no sizable corrections of this nature are found, see also 
Section \ref{sec:density_repulsive} and Ref. \cite{Mistakidis_orth_cat}.

\subsection{Entanglement Dynamics}\label{sec:entanglement_attractive}

\begin{figure*}[ht]
 	\centering
  	\includegraphics[width=1.0\textwidth]{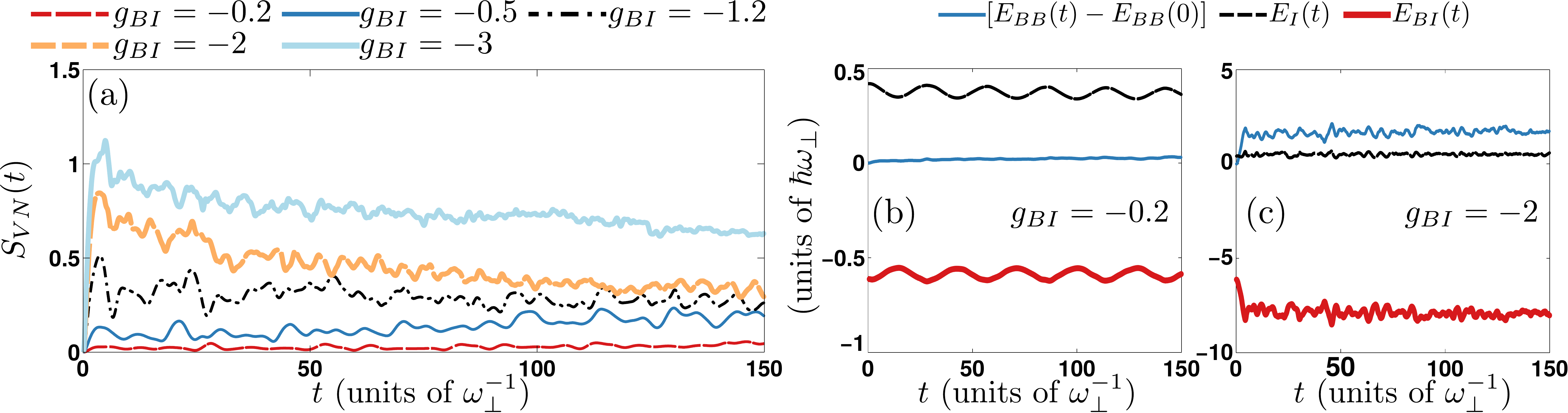}
     \caption{(a) Time-evolution of the Von-Neumann entropy for different attractive interspecies interaction strengths (see legend).  
     (b) Expectation value of the energy of the bosonic bath $E_{B}(t)$, the impurity $E_{I}(t)$ and their interspecies interaction energy $E_{BI}(t)$ following an 
     interspecies interaction quench from $g_{BI}=0$ to (b) $g_{BI}=-0.2$ and (c) $g_{BI}=-2$. 
     Note the different energy scales of (b) versus (c). 
     The remaining system parameters are the same as in Fig. \ref{fig:8}.} 
     \label{fig:11}
\end{figure*}

To reveal the correlated character and in particular the degree of entanglement of the quench-induced dynamics we calculate the corresponding 
Von-Neumann entropy $S_{VN}(t)$ [see Eq. (\ref{eq:entropy})]. 
The time-evolution of $S_{VN}(t)$ is demonstrated in Fig. \ref{fig:11} (a) for different postquench interaction strengths $g_{BI}$. 
As in Section \ref{sec:entanglement_repulsive}, we again observe that $S_{VN}(t=0)=0$ holds for all cases due to the fact that initially $g_{BI}=0$. 
However for $t>0$ $S_{VN}(t)\neq 0$ testifying that the many-body state [Eq. (\ref{Eq:WF})] is entangled. 
At the initial stages of the dynamics, e.g. $0<t<5$ for $g_{BI}=-1.2$, $S_{VN}(t)$ exhibits its larger growth rate and subsequently shows an overall 
decreasing behavior tending to approach a constant value for large evolution times $t>120$. 
The fact that $S_{VN}(t)$ exhibits the aforementioned decreasing trend for $t>8$ can be explained via inspecting the 
dynamics of the underlying Schmidt coefficients $\lambda_k(t)$ of the many-body wavefunction [Eq. (\ref{Eq:WF})] (not shown here for brevity).  
At the early stages of the dynamics $\lambda_1(t)$ drops from unity very quickly, while $\lambda_2(t)$, $\lambda_3(t)$ (with $\lambda_2(t)>\lambda_3(t)$) acquire  
finite values which become maximal at the time-instant where the impurity emits a small portion of its  
density, see the dashed rectangle in Fig. \ref{fig:7} (f) at $t\approx4$. 
Thereafter, the central density hump of the impurity is predominantly described by $\abs{\Psi_1^I(x;t)}^2$ while the emitted density portion by a superposition of 
$\abs{\Psi_2^I(x;t)}^2$ and $\abs{\Psi_3^I(x;t)}^2$.  
For later evolution times the emitted density portion re-collides with the central hump [see the dashed rectangle in Fig. \ref{fig:7} (f) around $t\approx 20$]. 
Simultaneously $\lambda_1(t)$ tends to larger values, while the populations of $\lambda_2(t)$ and $\lambda_3(t)$ decrease. 
Then, $\abs{\Psi_1^I(x;t)}^2$ provides the dominant contribution to $\ket{\Psi_{MB}(t)}$. 
The above-described decrease of the higher-lying Schmidt coefficients leads to the decreasing tendency of $S_{VN}(t)$. 
This decreasing behavior of $S_{VN}$ is more pronounced for larger values of $g_{BI}$ and essentially indicates the attenuation of the oscillation amplitude 
of the impurity dictated in $\rho_I^{(1)}(x;t)$ and $\braket{X_I(t)}$, see for instance Figs. \ref{fig:7} (f) and Fig. \ref{fig:9}. 
It is worth mentioning here that the attenuation of $\braket{X_I(t)}$ is also related to an energy transfer from the impurity to the 
BEC (see Section \ref{sec:energy_attractive}). 
Therefore, the decreasing tendency of $S_{VN}(t)$ is reminiscent of a cooling process for the impurity atom. 
Moreover, by inspecting Fig. \ref{fig:11} (a) it becomes evident that entering deeper to the attractive regime of interactions leads to a larger magnitude of entropy, 
e.g. compare $S_{VN}(t)$ for $g_{BI}=-0.5$ and $g_{BI}=-2$. 

\subsection{Interspecies Energy Exchange}\label{sec:energy_attractive} 

To further comprehend the dissipative motion of the impurity through the BEC for attractive interspecies interactions we also investigate 
the dynamics of the individual energy contributions of the bosonic mixture. 
The resulting energy parts following a quench to weak attractions, e.g. $g_{BI}=-0.2$, are presented in Fig. \ref{fig:11} (b). 
We observe that the interspecies interaction energy, $E_{BI}(t)$, and the energy of the impurity, $E_I(t)$, oscillate out-of-phase with one another in time 
with a weak amplitude taking negative and positive values respectively. 
Also when $E_{BI}(t)$ is maximized the corresponding $E_I(t)$ minimizes since then the impurity resides in regions of lower BEC density. 
The reverse process occurs when the impurity is close to the trap center, i.e. $E_{BI}(t)$ minimizes and accordingly $E_{I}(t)$ is maximized. 
On the other hand, the energy of the BEC $E_{B}(t)$ shows a minor increase at the very early stages of the dynamics and subsequently it remains constant. 
This increasing tendency of $E_{B}(t)$ indicates that the impurity conveys a minor amount of energy to the BEC. 

Turning to strong attractive interspecies interactions, e.g. $g_{BI}=-2$ demonstrated in Fig. \ref{fig:11} (c) the energy contributions 
exhibit a completely different behavior. 
At the very early stages of the dynamics, i.e. $0<t<5$, the energy of the bath $E_{B}(t)$ and the impurity $E_{I}(t)$ increase whilst the interaction energy 
$E_{BI}(t)$ reduces.
The increasing behavior of $E_{I}(t)$ indicates that the impurity gains kinetic energy due to the quench transferring also a part of its energy to the bosonic 
gas \cite{Mistakidis_orth_cat,Nielsen} which creates sound waves, see also Fig. \ref{fig:7} (f). 
For later evolution times $E_{I}(t)$ remains almost constant since the impurity is strongly localized while $E_{B}(t)$ and $E_{BI}(t)$ fluctuate due to the 
existence of sound waves in the BEC background \cite{Katsimiga_diss_flow}.

\section{Effective Mass}\label{sec:effective_mass}

Having analyzed the nonequilibrium dynamics of the subsonic impurity which penetrates the BEC we next measure its effective mass $m^{eff}$. 
We remark that the effective mass of quasiparticles has been measured experimentally based on the collective excitations of the impurities, e.g. their 
breathing motion \cite{Catani,Scazza}. 
Recall here that for weak repulsive interactions, $0<g_{BI}<0.95$, the impurity moves back and forth with respect to the trap center and remains within the bosonic bath. 
Entering the strong repulsive regime, $g_{BI}>1$, it probes the left edge of the BEC where it equilibrates for longer evolution times. 
However for attractive interspecies coupling strengths it performs a damped oscillatory motion within the bosonic medium. 
In all cases, since the impurity interacts with the BEC it is dressed by the excitations of the latter forming a quasiparticle. 
To model the motion of the impurity within the BEC we assume that it follows the following effective damped equation of motion   
\begin{equation}
  \ddot{x}+\frac{\gamma^{eff}}{m^{eff}}\dot{x}=-\frac{\omega^{eff}}{m^{eff}}x.\label{effective_equation}
\end{equation} 
Here, $\omega^{eff}$ refers to the effective trapping of the formed quasiparticle due to the combined effect of its interaction with the bath and 
the presence of the external harmonic confinement. 
Furthermore, $m^{eff}$ denotes the effective mass of the impurity and $\gamma^{eff}$ is the effective damping parameter of the impurity due to its motion 
inside the BEC. 
We also remark that the above effective description inherently involves the assumption that the impurity is effectively trapped by the bosonic bath. 
Therefore it is valid only for the interaction interval $-2.5<g_{BI}<0.95$ where the impurity does not escape from the Thomas-Fermi radius of the BEC. 

\begin{figure*}[ht]
 	\centering
  	\includegraphics[width=1.0\textwidth]{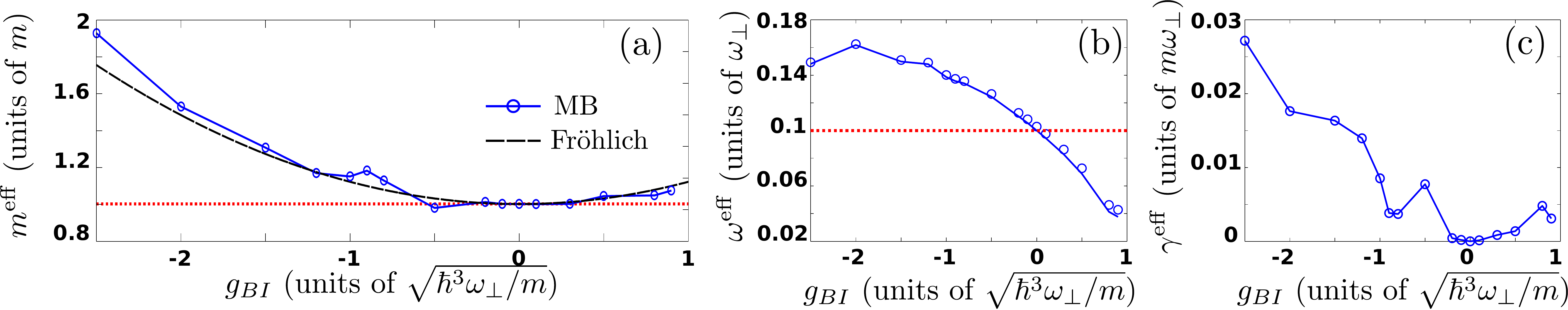}
     \caption{(a) Effective mass of the quasiparticle for different postquench interspecies interaction strengths $g_{BI}$ calculated within the 
     Fr\"ohlich and the many-body (MB) approach (see legend).  
     (b) Effective trapping frequency of the quasiparticle and (c) damping parameter of its motion into the BEC background for varying $g_{BI}$. 
     The dashed lines in (a), (b) indicate the bare value of the depicted quantity. 
     In all cases $N_{B}=100$, $N_I=1$, $g_{BB}=1$ and the frequency of the external harmonic confinement is $\omega=1$. 
     Initially the impurity is non-interacting with the bosonic bath and moves with a velocity $u_0=-u_c/2$. 
     The quench is performed when the impurity is located at $x_0=0$.} 
     \label{fig:12}
\end{figure*}

To determine the effective mass of the formed quasiparticle as well as its effective trapping frequency and damping parameter within $-2.5<g_{BI}<0.95$, 
we perform the following analysis. 
We first calculate the mean position, $\braket{X_I(t)}$, and momentum, $\braket{P_I(t)}$, of the impurity for a fixed interspecies interaction quench solely 
relying on our numerical calculations described in Sections \ref{sec:quench_repulsive} and \ref{sec:quench_attractive}. 
Then, by solving Eq. (\ref{effective_equation}) it can be easily shown that the mean position of the impurity reads   
\begin{equation}
\begin{split}
\braket{X_I(t)}=e^{-\frac{\gamma^{eff}}{2m^{eff}}t}&\Big[x_0 \cos(\omega_{0}t)\\&-\frac{p_0+\frac{\gamma^{eff}}{2}x_0}{m^{eff}\omega_0}\sin(\omega_0t)\Big], \label{position_theory}
\end{split}
\end{equation} 
with $\omega_0=\sqrt{(\omega^{eff})^2-(\frac{\gamma^{eff}}{2})^2}$. 
Moreover, since we consider that initially the impurity-BEC interaction is zero, i.e. $g_{BI}=0$, we obtain 
$p_0\equiv\braket{\Psi(0)|\hat{p}|\Psi(0)}= \hbar m u_0$ and $x_0\equiv\braket{\Psi(0)|\hat{x}|\Psi(0)}=0$. 
Also, the mean momentum of the impurity obeys the following equation  
\begin{equation}
\begin{split}
 &\braket{P_I(t)}=e^{-\frac{\gamma^{eff}}{2m^{eff}}t}\bigg\{ p_0 \cos(\omega_{0}t)\\&+\Big[m^{eff}\omega_0x_0+\frac{\gamma^{eff}(p_0+\frac{\gamma^{eff}}{2}x_0)}{2m^{eff}\omega_0}\Big]\sin(\omega_0t) \bigg\}. \label{momentum_theory}
\end{split}
\end{equation}
Evidently, in the above equations the unknown parameters that need to be determined are $\omega^{eff}$, $m^{eff}$ and $\gamma^{eff}$. 
In order to estimate these parameters we perform a fitting of the analytical form of both $\braket{X_I(t)}$ and $\braket{P_I(t)}$ provided 
by Eqs. (\ref{position_theory}) and (\ref{momentum_theory}) to the corresponding numerically obtained results of $\braket{X_I(t)}$ and $\braket{P_I(t)}$. 
The values of the parameters $\omega^{eff}$, $m^{eff}$ and $\gamma^{eff}$ obtained via this fitting procedure are shown in Fig. \ref{fig:12} for a varying 
$g_{BI}$ such that $-2.5<g_{BI}<0.95$ where the quasiparticle picture is well defined. 

Focusing on the attractive regime of interactions we observe that the effective mass of the emergent quasiparticle is larger than its bare mass and 
tends to the latter, i.e. $m^{eff}\to m$, as the non-interacting limit is approached. 
Additionally, the effective trapping frequency of the quasiparticle is larger than the actual frequency of the external harmonic oscillator and 
overall $\omega^{eff}$ exhibits a decreasing tendency as $g_{BI}\to0$. 
This result is in line with the previously discussed effective potential picture $V_I^{eff}(x)$, see Section  \ref{sec:density_attractive} 
and Fig. \ref{fig:10} (b). 
Moreover, the effective damping parameter $\gamma^{eff}$ acquires a finite value signaling the dissipative motion of the impurity inside the 
BEC and it tends to vanish for $g_{BI}\to 0$. 
Turning to repulsive interactions the quasiparticle effective mass is very close to the bare value, e.g. $m^{eff}\approx 1.001 m$ at $g_{BI}=0.1$, while 
for increasing repulsion it becomes slightly larger, namely $m^{eff}\approx 1.043 m$ at $g_{BI}=0.5$. 
Note here that this behavior of $m^{eff}$ in the repulsive regime of interactions is in contrast to the one discussed in Ref. \cite{Mistakidis_eff_mass} 
where $m^{eff}$ has been found to become smaller than the bare mass of the impurity. 
In this latter case the effective potential used to describe the quasiparticle formation did not include a damping parameter, an assumption which has been 
proved sufficient for the zero velocity impurity. 
However in the present case the damping term is important for the description of the observed dynamics and it is responsible for the aforementioned discrepancy. 
Also, the effective trapping frequency is smaller than the one of the external harmonic oscillator and shows a decreasing tendency for larger repulsions. 
This behavior is in accordance with the effective potential picture introduced in Section \ref{sec:density_repulsive}, see also Fig. \ref{fig:4} and 
Eq. (\ref{effective_potential_repulsive}). 
Furthermore, $\gamma^{eff}$ takes small values and increases slightly as $g_{BI}$ becomes stronger. 

To expose the role of non-perturbative effects in the resulting effective mass of the quasiparticle we further calculate $m^{eff}$ relying 
on the well-known perturbative expansion of the Fr\"ohlich model \cite{Grusdt_approaches}. 
Note that this model operates in the absence of an external confinement, i.e. $\omega=0$. 
Indeed, it can be shown that the leading order correction of the effective mass \cite{Grusdt_approaches,Devreese_notes,Grusdt_1D} 
with respect to $g_{BI}$ is given by 
\begin{equation}
 m^{eff}=m_I+4g_{BI}^2A+ \mathcal{O}(g_{BI}^4).\label{frohlich} 
\end{equation}
In this expression, the constant $A=\int_0^{\infty}dk \frac{k^2(V_k/g_{BI})^2}{(\omega_k+k^2/2m_{I})^3}$ with the scattering amplitude defined by  
$V_k=\sqrt{n_0}(2\pi)^{-1/2}g_{BI}(\frac{(\xi k)^2}{2+(\xi k)^2})^{1/4}$. 
Also the healing length and the speed of sound of the BEC are $\xi=(2m_Bg_{BB}n_0)^{-1/2}$ and $u_c=\sqrt{\frac{g_{BB}n_0}{m_B}}$ respectively 
with $n_0$ being the density of the homogeneous bosonic gas. 
To adequately compare the prediction given by Eq. (\ref{frohlich}) with our results which include an external trap we choose $n_0=\rho^{(1)}_B(x=0;t=0)$.  
Furthermore, the dispersion relation of the elementary excitations of the bosonic gas corresponds to $\omega_k=u_ck \sqrt{1+\frac{1}{2}(k \xi)^2}$. 
Figure \ref{fig:12} (a) shows $m^{eff}$ within the Fr\"ohlich model for varying $g_{BI}$. 
Strikingly, the predictions of the Fr\"ohlich model and the full many-body approach are in very good agreement with one another 
both at weak attractive and repulsive interspecies interactions. 
Therefore we can deduce that for such weak interspecies interaction strengths the external trapping does not play any crucial role 
for the effective mass of the impurity. 
This result is not surprising since at these weak interactions the impurity resides well inside the bosonic gas and thus approximately 
experiences a homogeneous background. 
Noticeable deviations are observed for strong attractive interactions $g_{BI}>-1.25$, e.g. being of the order of $3\%$ and $9\%$ 
at $g_{BI}=-2.0$ and $g_{BI}=-2.5$ respectively. 
Also, small differences on $m^{eff}$ estimated between the two aforementioned approaches occur on the repulsive regime of interactions, and 
especially for $g_{BI}\geq 0.8$ become larger than $3\%$. 
These deviations can be partly attributed to the effect of the trap since for an increasing repulsion the density of the impurity probes the 
edges of the cloud of the BEC. 
Similarly the effect of the harmonic trap cannot be neglected for large attractive interactions. 
Indeed the effective potential in this case amplifies any small discrepancies of the BEC density from the homogeneous 
case that occur around the trap center. 

As a final remark we note that the effective mass depends weakly on the initial, i.e. before the quench, velocity $u_0$ of the impurity. 
For instance, referring to a fixed postquench interspecies interaction strength e.g. $g_{BI}=-1$ the effective mass takes values 
$m^{eff}\approx1.19$ for $u_0\approx-u_c$, $m^{eff}\approx1.15$ when $u_0=-u_c/2$ and $m^{eff}\approx1.13$ if $u_0=-u_c/5$. 

\section{Single-Shot Simulations}\label{sec:single_shots}

To provide further possible experimental links of our results we simulate in-situ single-shot absorption 
measurements \cite{Sakmann_single_shot,mistakidis_phase_sep} aiming at demonstrating how in-situ imaging can 
be used to adequately monitor the quench-induced dynamical dressing of the impurity. 
These measurements probe the spatial configuration of the atoms and therefore the many-body probability distribution 
which is indeed available within ML-MCTDHX. 
The corresponding experimental images are obtained via a convolution of the spatial particle configuration with a point spread function 
that essentially dictates the experimental spatial resolution. 
Below, we present such simulations by employing a point spread function of Gaussian shape and width $w_{PSF}=1 \ll l\approx 3.16$, 
with $l=\sqrt{1/\omega}$ being the harmonic oscillator length. 
Note also that $w_{PSF} > \xi \approx 0.4$, where $\xi=\frac{\hbar}{\sqrt{2}mu_c}$ denotes the healing length of the BEC. 
For a more elaborated discussion on the details of the numerical implementation of this process in one-dimensional binary systems we refer 
the reader to Appendix \ref{sec:single_shot_algorithm} and also to Refs. \cite{mistakidis_phase_sep,Erdmann_phase_sep}. 

\begin{figure*}[ht]
 	\centering
  	\includegraphics[width=1.0\textwidth]{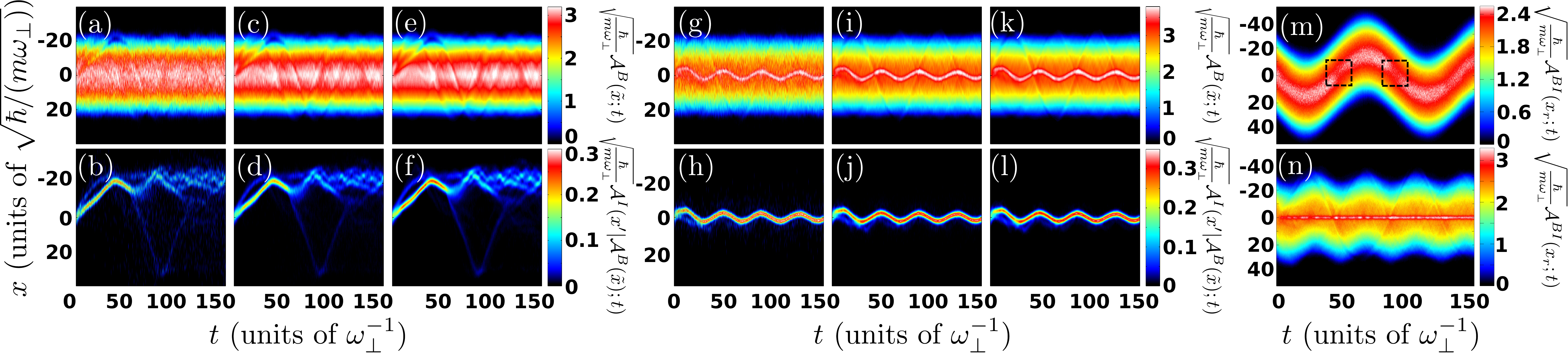}
     \caption{Averaged images of each species over $N_{shots}=10$ (left panels), $N_{shots}=100$ (central panels) and $N_{shots}=800$ (right panels) following an 
     interspecies interaction quench from $g_{BI}=0$ to (a)-(f) $g_{BI}=1.5$ and (g)-(l) $g_{BI}=-2$. 
     Average images over $N_{shots}=800$ of the bosonic gas in the co-moving frame of the impurity when quenching the interspecies interaction strength 
     from $g_{BI}=0$ to (m) $g_{BI}=0.5$ and (n) $g_{BI}=-2$.   
     The remaining system parameters are the same as in Fig. \ref{fig:8}.} 
     \label{fig:13}
\end{figure*} 

Having at hand the many-body wavefunction of our system within ML-MCTDHX we reproduce in-situ single-shot images for the BEC medium 
$B$, $\mathcal{A}^B(\tilde{x};t)$, and the impurity $I$, $\mathcal{A}^I(\tilde{x'}|\mathcal{A}^B(\tilde{x});t_{im})$, 
at each time-instant of the evolution. 
Here, $t_{im}$ denotes the time-instant of the imaging. 
In particular, we consecutively image first the BEC and then the impurity species. 
Note that the reverse imaging process does not affect the image obtained after averaging over several single-shots, see also Appendix \ref{sec:single_shot_algorithm}. 
Further details of the corresponding simulation process of this experimental technique are discussed in Appendix \ref{sec:single_shot_algorithm}. 
In the following we analyze the nonequilibrium dynamics of the bosonic mixture for quenches towards strongly repulsive, $g_{BI}=1.5$, and 
attractive interactions, i.e. $g_{BI}=-2$. 
We remark that a similar analysis has been followed also for other values of $g_{BI}$ (not shown here for brevity reasons). 
Before describing the outcome of the images it is noteworthy to mention that a direct correspondence between the single-particle density 
and only one single-shot image is not possible due to the small particle number of the considered setup, $N_B=100$ and $N_I=1$. 
Such a resemblance is feasible only when considering large particle numbers, e.g. of the of order $10^5$ particles \cite{Katsimiga_diss_flow}. 
Another reason that excludes the possibility of explicitly observing the one-body density within a single-shot image is the 
presence of multiple orbitals in the system  [Eqs. (\ref{Eq:SPF_bath}) and (\ref{Eq:SPF_impurity})]. 
More specifically, the many-body state is expressed as a superposition over multiple orbitals [see Eqs. (\ref{Eq:SPF_bath}) and (\ref{Eq:SPF_impurity})] 
and thus imaging an atom alters the many-body state of the other atoms and as a consequence their one-body density. 
A more elaborated discussion on this topic is provided in Refs. \cite{mistakidis_phase_sep,Katsimiga_bent,Katsimiga_diss_flow}. 
Most importantly, it can be demonstrated that the average image e.g. of the BEC ($B$ species) i.e. $\mathcal{\bar{A}}^{B}(\tilde{x})$, 
over a sample of $N_{shots}$ single-shot images, $\mathcal{A}^{B}(\tilde{x})$, is related to the $B$ species one-body density, 
namely $\rho_{B}^{(1)} (x_{B}')$, as follows   
\begin{equation}
 \mathcal{\bar{A}}^{B}(\tilde{x})=\frac{N_{B}}{\sqrt{2\pi}w_{PSF}}\int dx_{B}' e^{-\frac{(\tilde{x}-x_{B}')^2}{2w^2_{PSF}}} \rho_{B}^{(1)} (x_{B}').
 \label{average_images} 
\end{equation}
In this expression, $\tilde{x}$ are the spatial coordinates within the image and $x_{B}'$ refers to the spatial coordinate of the $B$ species.   
The width of the employed point spread function is $w_{PSF}$ and the species particle number is $N_{B}$. 
A similar relation holds for the other species but using the corresponding images. 

According to our above discussion in order to retrieve the one-body density of each subsystem we rely on an average of several single-shot images 
for each species. 
In particular, we measure $\bar{\mathcal{A}}^B(\tilde{x};t)=1/N_{shots}\sum_{k=1}^{N_{shots}} 
\mathcal{A}_k^B(\tilde{x};t)$ for the BEC and $\bar{\mathcal{A}}^I(\tilde{x}^{'}|\mathcal{A}^B(\tilde{x});t)=1/N_{shots}\sum_{k=1}^{N_{shots}} 
\mathcal{A}_k^I(\tilde{x}^{'}|\mathcal{A}^B(\tilde{x});t)$ for the impurity atom respectively. 
Figures \ref{fig:13} (a)-(f) show $\bar{\mathcal{A}}^B(\tilde{x};t)$ and 
$\bar{\mathcal{A}}^I(\tilde{x}^{'}|\mathcal{A}^B(\tilde{x});t)$ for different number of samplings, i.e. $N_{shots}$, upon considering a 
quench from $g_{BI}=0$ to $g_{BI}=1.5$. 
Comparing this averaging process for an increasing number of $N_{shots}$ and the actual single-particle density calculated via 
ML-MCTDHX [see Figs. \ref{fig:2} (e), (f)] unveils that they become almost the same. 
More specifically, it can easily be deduced that for $N_{shots}>100$ the 
$\bar{\mathcal{A}}^A(\tilde{x};t)$ and the $\bar{\mathcal{A}}^B(\tilde{x}^{'}|\mathcal{A}^A(\tilde{x});t)$ 
tend gradually to $\rho^{(1)}_{B}(x;t)$ and $\rho^{(1)}_{I}(x;t)$ respectively. 
The same overall phenomenology occurs for the case of a quench towards the attractive interaction regime, here $g_{BI}=-2$, as 
illustrated in Figs. \ref{fig:13} (g)-(k). 
Indeed, the dissipative motion of the impurity and its imprint on the BEC background are fairly captured even for $N_{shots}=10$, 
e.g. compare Figs. \ref{fig:13} (g)-(h) with Figs. \ref{fig:7} (e)-(f). 

Utilizing the aforementioned single-shots we can further probe the spatial configuration of the bosonic gas in the co-moving frame of the impurity. 
This procedure sheds light on the imprint of the impurity motion onto the correlations emanating within the bosonic mixture. 
Such a protocol has been succesfully experimentally implemented to probe the internal structure of magnetic polarons \cite{Koepsell}. 
Within this protocol we shift each of the previously obtained single-shots, $\mathcal{A}_k^{B}(\tilde{x},t)$
by the amount $X^I_k=\int d\tilde{x}' ~ \tilde{x}' \mathcal{A}_k^I(\tilde{x}^{'}|\mathcal{A}^B(\tilde{x});t)$ being the measured position of the
impurity at the $k$-th single-shot, i.e. $\mathcal{A}_k^{BI}(x_r,t)=\mathcal{A}_k^{B}(\tilde{x}-X^I_k,t)$. 
It can be shown that the corresponding average image $\mathcal{\bar{A}}^{BI}(x_r,t)=\sum_{i=1}^{N_{shots}} \mathcal{A}_k^{BI}(x_r,t)$ over 
$N_{shots}$ is related to the two-body interspecies correlation function as follows
\begin{equation}
\begin{split}
\mathcal{\bar{A}}^{BI}(x_r,t)=\frac{N_B}{\sqrt{2\pi}w_{PSF}}&\int dx'_{B} d x'_I e^{-\frac{[x_r-(x_{B}'-x'_I)]^2}{2w^2_{PSF}}} \\& \times \rho_{BI}^{(2)} (x'_{B}, x'_I;t).
\end{split}
\end{equation} 
Here, $\rho_{BI}^{(2)}(x_1,x_2;t)=\langle\Psi_{MB}(t)|\hat\Psi^{\dagger B}(x_1)\hat\Psi^{\dagger I}(x_2)\hat\Psi^{B}(x_1)$ $\hat\Psi^{ I}(x_2)|\Psi_{MB}(t) \rangle$ are  
the diagonal elements of the two-body interspecies reduced density matrix \cite{mistakidis_phase_sep,Sakmann_den_matr}.
The latter provides the probability of measuring a $B$- and a $I$-species particle simultaneously at positions $x_1$ and $x_2$ respectively. 
$\bar{A}^{BI}(x_r,t)$ obtained from $N_{shots}=800$ is presented in Figs. \ref{fig:13} (m) and (n) exemplarily for a quench to positive $g_{BI}=0.5$ 
and negative $g_{BI}=-2$ interactions respectively. 
Recall that the effective quasiparticle description holds only when the impurity is effectively trapped into the bosonic bath and therefore 
in our case is valid for $-2.5<g_{BI}<0.95$. 
For repulsive postquench interactions we observe that $\bar{A}^{BI}(x_r,t)$ shows an overall oscillatory behavior which is a consequence of the mere 
fact that the impurity undergoes for these interactions an oscillatory motion inside the bosonic gas, see also Fig. \ref{fig:2} (b). 
Focusing on $\bar{A}^{BI}(x_r,t)$ in the vicinity of the impurity, i.e. $x_r \approx 0$, we deduce that the latter repels the particles of the bosonic gas leading
to the development of shallow dips in $\bar{A}^{BI}(x_r,t)$, 
e.g. see the dashed rectangles in Fig. \ref{fig:13} (m). 
Turning to strong attractive interactions, see Fig. \ref{fig:13} (n), we discern the formation of a pronounced peak in the vicinity of $x_r=0$ caused by the presence of the 
impurity [see also Fig. \ref{fig:7} (f)]. 
This result is in accordance to the effective potential picture [Eq. (\ref{effective_pot_bath}) and Fig. \ref{fig:10}] 
Also the height of this central peak fluctuates which is hardly discernible in Fig.  \ref{fig:13} (n). 
Additionally, an overall oscillatory behavior of $\bar{A}^{BI}(x_r,t)$ occurs since we operate in the co-moving frame of the impurity. 
Concluding based on $\bar{A}^{BI}(x_r,t)$  we deduce that the interspecies two-body correlations between the impurity and the BEC are 
much more prevalent in the case of attractive interactions.

\section{Summary and Conclusions}\label{sec:conclusions} 

We have studied the interspecies interaction quench quantum dynamics of a subsonically moving impurity that penetrates a 
harmonically trapped Bose-Einstein condensate. 
Monitoring the time-evolution of the impurity on the single-particle level we identify a variety of response 
regimes arising for different interaction strengths. 

For weak postquench interspecies repulsive interactions the subsonic impurity performs a dipole motion inside the bosonic bath. 
The latter remains essentially unperturbed exhibiting some small distortions from its initial Thomas-Fermi profile. 
Increasing the interspecies coupling, the oscillation period of the impurity becomes larger and shallow density dips built upon the 
bosonic density thus imprinting the impuritys' motion. 
However, at strong quench amplitudes such that the interspecies interaction exceeds the bosonic intraspecies one the dynamical 
behavior of the impurity is significantly altered. 
More specifically, the impurity travels in the direction of its initial velocity towards the corresponding edge of the BEC background 
and thereafter fluctuates around the Thomas-Fermi radius. 
This latter behavior of the impurity at strong repulsive interactions occurs independently of its initial velocity, its prequench position, 
the trapping frequency and the mass ratio of the atomic species. 
Most importantly it takes place due to the involvement of correlations since e.g. within the mean-field approximation the impurity undergoes an oscillatory 
motion inside the bosonic bath throughout the dynamics. 
Employing the Von-Neumann entropy reveals the development of strong interparticle correlations in the course of the evolution, a result that becomes more pronounced 
for larger repulsions. 
Inspecting the individual energy contributions of each species we unveil that the impurity dissipates energy into the bosonic medium, a phenomenon that is 
more enhanced for increasing interspecies interactions. 
To interpret the dynamics of the impurity we construct an effective potential which corresponds to a modified harmonic oscillator for weak interactions 
turning to a double-well when entering the strongly repulsive regime. 

Entering attractive interspecies interactions we showcase that the impurity undergoes a damped oscillatory motion inside the bosonic bath. 
This behavior becomes more pronounced for an increasing attraction where the impurity exhibits a localization tendency and the BEC develops 
a density peak at the location of the impurity. 
It is shown that the above response of each species can be intuitively understood in terms of an effective potential picture for the bath and the 
impurity independently. 
Moreover, by invoking the energy contributions of each species we find that the impurity transfers a part of its energy to the bosonic medium 
which in turn generates sound waves being also evident in its single-particle density. 
Also, an inspection of the Von-Neumann entropy shows the presence of interspecies correlations especially for stronger attractive interactions. 

We have estimated the effective mass of the emergent quasiparticle by modeling its damped motion through the medium with an effective dissipative equation of motion.  
Performing a fitting of our numerical results and the analytical prediction of this dissipative equation we are able to estimate the effective mass, trapping 
frequency and damping parameter of the impurity. 
It is found that in the attractive regime of interactions the effective mass and trapping frequency are larger than the bare ones and tend 
to the latter when approaching the non-interacting limit. 
Also, the effective damping parameter acquires a finite value and tends to vanish for zero interspecies couplings. 
For repulsive interactions the quasiparticles' effective mass is slightly larger than its bare value while the damping parameter acquires small values. 
The corresponding effective trapping frequency is smaller than the one of the external harmonic oscillator showing a decreasing tendency 
for larger repulsions. 
Finally, we have provided possible experimental evidences of the impurity dynamics by simulating in-situ single-shot measurements. 
In particular, we showcase how an increasing sampling of such images can be used to adequately retrieve the observed dynamics. 

There is a variety of possible extensions of the present work in future endeavors. 
An imperative prospect is to unravel the resultant interspecies interaction quench dynamics upon considering two or more interacting bosonic 
impurities immersed in a bosonic bath. 
This study will shed light into the presence of the most probably emergent induced interactions between the impurities and would enable us to 
systematically explore their role in the time-evolution.  
Additionally, the inclusion of temperature effects in such an investigation would be very interesting \cite{Tajima,Liu}. 
Another intriguing direction would be to simulate the corresponding radiofrequency spectrum \cite{Mistakidis_Fermi_pol} or the structure 
factor of the current setup \cite{Mistakidis_orth_cat,Shchadilova} by employing spinor impurities in order to identify the possibly emerging polaronic states and 
subsequently measure e.g. their lifetime and residue. 
Certainly the generalization of the present results to higher-dimensional settings is highly desirable.

\appendix

\section{Technical Details of the Single-Shot Algorithm} \label{sec:single_shot_algorithm}

To perform the simulation of the single-shot procedure we employ a sampling of the many-body probability 
distribution \cite{Sakmann_single_shot,Katsimiga_diss_flow,mistakidis_phase_sep}. 
The latter is available in terms of the ML-MCTDHX approach for each time-instant of the evolution. 
It is important to note at this point that the numerical implementation of this experimental procedure has already 
been reported and applied to a variety of setups including neutral and spinor atoms \cite{Katsimiga_bent,Katsimiga_diss_flow,Koutentakis_prob} 
as well as binary mixtures \cite{mistakidis_phase_sep,Erdmann_phase_sep}. 
In this sense, below we briefly outline the corresponding numerical procedure but for more details and extensive discussions we 
refer the reader to Refs. \cite{Sakmann_single_shot,mistakidis_phase_sep,Katsimiga_bent,Katsimiga_diss_flow}. 

As it has already been argued in previous works \cite{Sakmann_single_shot,mistakidis_phase_sep,Katsimiga_bent,Katsimiga_diss_flow,Koutentakis_prob}, 
the single-shot procedure for binary mixtures is crucially affected by the systems' intra- and interspecies correlations. 
Indeed, for a many-body state the presence of entanglement [see Eq. (\ref{Eq:WF})] among the distinct species is important 
regarding the image ordering. 
This dependence can be understood by resorting to the underlying Schmidt decomposition [see Eq. \ref{Eq:WF}] since it directly affects 
the Schmidt coefficients $\lambda_k$. 
Below, we briefly sketch the numerical process when imaging first the BEC $B$ and subsequently the impurity $I$ species. 
In this way, we obtain the corresponding absorption images $\mathcal{A}^B(\tilde{x})$ and $\mathcal{A}^I(\tilde{x}'|\mathcal{A}^B(\tilde{x}))$. 
To avoid any confusion, let us remark that in order to image first the $I$ and then the $B$ species we can follow the same procedure, 
retrieving the images $\mathcal{A}^I(\tilde{x})$ and $\mathcal{A}^B(\tilde{x}'|\mathcal{A}^I(\tilde{x}))$. 
It is also worth mentioning that the image ordering plays a role when one is interested in the individual single-shot images. 
However, in our case that we discuss the average of a sample of single-shots (see the discussion in Section \ref{sec:single_shots}) 
the image ordering is irrelevant since all the effects stemming from entanglement are averaged out. 

To perform the imaging of the $B$ and then of the $I$ species we first annihilate one-by-one all $B$-species bosons. 
Referring to a specific time-instant of the imaging, e.g. $t_{im}$, a random position is drawn obeying 
$\rho_{N_B}^{(1)}(x_1')>z_1$ where $z_1$ is a random number taking values in the interval 
[$0$, $ \max\lbrace{\rho^{(1)}_{N_B}(x;t_{im})\rbrace}$]. 
Then by utilizing the projection operator $\frac{1}{\mathcal{N}}(\hat{\Psi}^B(x_1')\otimes \hat{\mathbb{I}}_I)$ we project 
the ($N_B+N_I$)-body wavefunction onto the ($N_B-1+N_I$)-body one. 
The bosonic field operator annihilating a $B$ species boson at position $x_1'$ is $\hat{\Psi}^B(x_1')$ and 
$\mathcal{N}$ denotes the normalization constant. 
Evidently, this process affects the Schmidt coefficients, $\lambda_k$, and consequently the densities $\rho^{(1)}_{N_B-1}(t_{im})$ 
and $\rho^{(1)}_{N_I}(t_{im})$ are altered. 
As a result, the Schmidt decomposition of the many-body wavefunction following this first measurement reads
\begin{equation}
\begin{split}
&\ket{\tilde{\Psi}_{MB}^{N_B-1,N_I}(t_{im})}=\\ &\sum_i \sqrt{\tilde{\lambda}_{i,N_B-1}(t_{im})}\ket{\tilde{\Psi}_{i,N_B-1}^B(t_{im})}\ket{\Psi_i^I(t_{im})}.   
\label{Eq:wfn_first_measurement}
\end{split}
\end{equation} 
In this expression, $\ket{\tilde{\Psi}_{i,N_B-1}^B}=\frac{1}{N_i}\hat{\Psi}^B(x_1')\ket{\Psi_i^B}$ is the $N_B-1$ species 
wavefunction with $N_i=\sqrt{\bra{\Psi_i^B}\hat{\Psi}^{B\dagger}(x_1')\hat{\Psi}^B(x_1')\ket{\Psi_i^B}}$. 
Moreover, the Schmidt coefficients of the ($N_B-1+N_I$)-body wavefunction are given by $\tilde{\lambda}_{i,N_B-1}=\lambda_i N_i/\sum_i \lambda_i N_i^2$. 

The imaging process of the $B$-species is finalized after repeating the above-mentioned steps $N_B-1$ times realizing the following distribution 
of positions ($x'_1$, $x'_2$,...,$x'_{N_B-1}$). 
This distribution is then convoluted with a point spread function resulting in the single-shot 
image of the $B$-species $\mathcal{A}^B(\tilde{x})=\frac{1}{\sqrt{2\pi}w_{PSF}}\sum_{i=1}^{N_B}e^{-\frac{(\tilde{x}-x'_i)^2}{2w_{PSF}^2}}$, 
where $\tilde{x}$ are the spatial coordinates within the image and $w_{PSF}$ is the width of the point spread function. 
After annihilating all $N_B$ atoms, the many-body wavefunction acquires the form  
\begin{equation}
\begin{split}
&\ket{\tilde{\Psi}_{MB}^{0,N_B}(t_{im})}=\\ &\ket{0^B} \otimes\sum_i \frac{\sqrt{\tilde{\lambda}_{i,1}(t_{im})}
\braket{x'_{N_B}|\Phi_{i,1}^B}}{\sum_j{\sqrt{\tilde{\lambda}_{j,1}(t_{im})|\braket{x'_{N_B}|\Phi_{j,1}^B}|^2}}}\ket{\Psi_i^I(t_{im})}.   
\label{Eq:A3}
\end{split}
\end{equation} 
Here, the single-particle orbital of the $j$-th mode is $\braket{x'_{N_B}|\Phi_{j,1}^B}\equiv\braket{0^B|\hat{\Psi}^B(x'_{N_B})|\Phi_{j,1}^B}$ and 
the second term in the cross product of the right-hand side ($\ket{\Psi_{MB}^{N_I}(t_{im})}$) denotes the impurity species wavefunction. 
The latter is a non-entangled single-particle wavefunction ($N_I=1$) and as a consequence the corresponding single-shot procedure of the $I$ species 
reduces to that of a single-species \cite{Sakmann_single_shot,Katsimiga_bent,Katsimiga_diss_flow}. 
Indeed, for an imaging time $t=t_{im}$, we measure $\rho^{(1)}_{N_I}(x;t_{im})$ from $\ket{\Psi^{N_I}_{MB}}\equiv \ket{\Psi(t_{im})}$ and draw a 
random position $x''_1$ satisfying $\rho^{(1)}_{N_I}(x''_1;t_{im})>z_2$. 
Here, $z_2$ is a random number bounded in the interval [$0$, $\rho^{(1)}_{N_I}(x;t_{im})$]. 
As a result, the $I$-species particle is annihilated at position $x''_1$ and this position is subsequently convoluted with a point spread function 
resulting to the single-shot image $\mathcal{A}^I(\tilde{x'}|\mathcal{A}^B(\tilde{x}))$.

\section{Convergence of the Many-Body Simulations} \label{sec:convergence_numerics}

To simulate the correlated nonequilibrium quantum dynamics of the considered binary mixture we resort to the 
Multi-Layer Multi-Configurational Time-Dependent Hartree Method for Atomic Mixtures (ML-MCTDHX) \cite{MLX,MLB1}, see also Section \ref{ML_ansatz}. 
It is a variational approach for solving the time-dependent many-body Schr{\"o}dinger equation of atomic mixtures consisting either 
of bosonic \cite{mistakidis_phase_sep,Mistakidis_eff_mass,Mistakidis_orth_cat} or fermionic 
\cite{Erdmann_phase_sep,Koutentakis_prob} components that might additionally include spin degrees of freedom \cite{Mistakidis_orth_cat,Mistakidis_Fermi_pol}. 
More specifically, this method relies on the expansion of the many-body wavefunction in terms of a time-dependent and variationally optimized basis. 
Such a treatment enables us to include all the important inter- and intraspecies correlations into our many-body ansatz utilizing a computationally 
feasible basis size. 
In this way, it allows us to span the relevant subspace of the Hilbert space at each time-instant in an efficient manner. 
The latter is in contrast to methods relying on a time-independent basis where the number of basis states can be significantly larger rendering 
the simulation of intermediate size systems impossible. 

The underlying Hilbert space truncation is inferred from the used orbital configuration space, namely $C=(D;d_B;d_I)$. 
In this notation, the number of species and single-particle functions of each species are denoted by $D=D_B=D_I$ and $d_B$, $d_I$ respectively, 
see also Eqs. (\ref{Eq:WF}), (\ref{Eq:SPF_bath}) and (\ref{Eq:SPF_impurity}). 
We remark here that since we use a single impurity then by definition $D=d_I$ holds. 
Additionally, for our numerical calculations a primitive basis based on a sine discrete variable representation 
consisting of 1000 grid points is employed. 
This sine discrete variable representation intrinsically introduces hard-wall boundary conditions which in our case 
are imposed at $x_\pm=\pm80$. 
Of course, the location of these boundaries do not affect our results since we do not observe appreciable densities to occur 
beyond $x_{\pm}=\pm40$. 
\begin{figure*}
  	\includegraphics[width=0.9\textwidth]{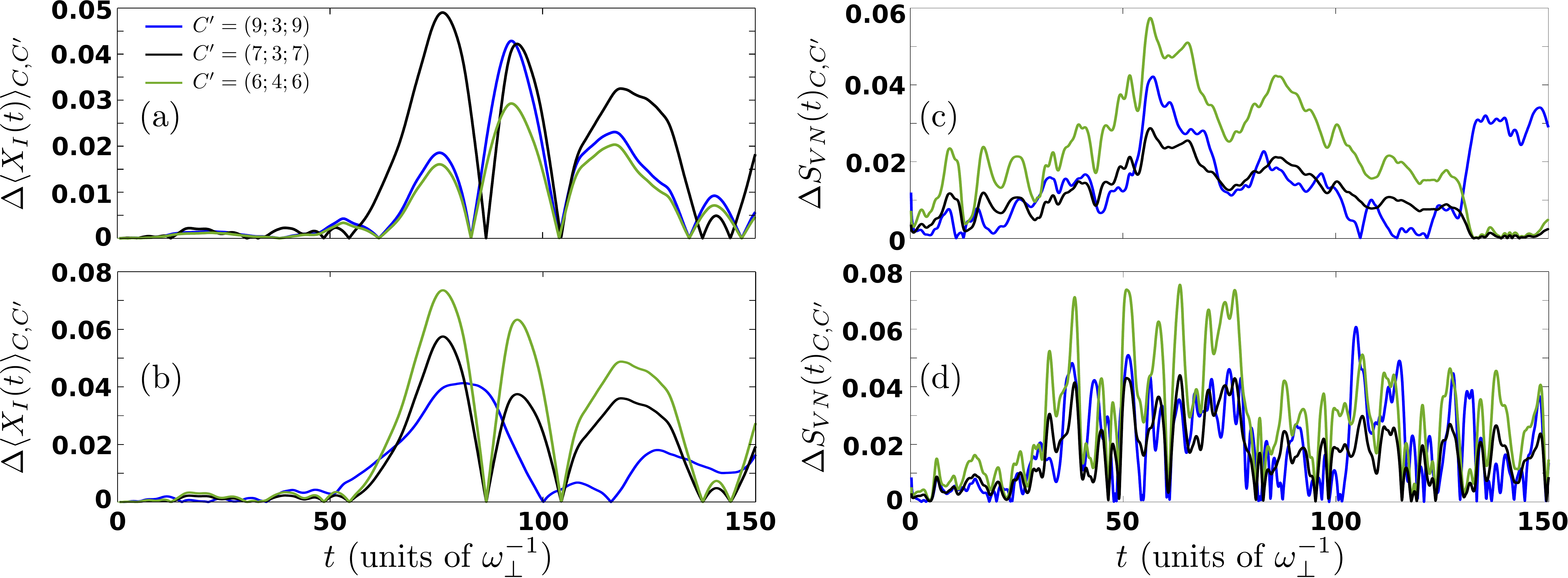}
  	\caption{Dynamics of the deviation (a), (b) $\Delta \braket{X_I(t)}_{C,C'}$ of the position of the impurity and (c), (d) $\Delta S_{VN}(t)_{C,C'}$ of 
  	the Von-Neumann entropy measured between the $C=(8;3;8)$ and other orbital configurations $C'=(D;d_B;d_I)$ (see legend).  
  	The system consists of $N_B=100$ bosons prepared in their ground state of $g_{BB}=1$ and $N_I=1$ impurity. 
  	Initially $g_{BI}=0$ and at $t=0$ we perform an interspecies interaction quench to (a), (c) $g_{BI}=2$ and (b), (d) $g_{BI}=-2$.}
  	\label{fig:14} 
  \end{figure*}

To conclude upon the convergence of our many-body simulations we ensure that all observables of interest become almost insensitive, to a certain  
degree, when varying the used orbital configuration space i.e. $C=(D;d_B;d_I)$. 
In our case, a convergent behavior of all the many-body calculations discussed in the main text has been achieved by exploiting the 
orbital configuration space $C=(8;3;8)$. 
To testify the convergence of our results for a different number of species and single-particle functions e.g. we examine the mean position 
of the impurity during the interspecies interaction quench dynamics. 
In particular, we investigate its absolute deviation between the $C=(8;3;8)$ and other orbitals configurations $C'=(D;d_B;d_I)$ 
\begin{equation}
\Delta \braket{X_I(t)}_{C,C'} =\frac{\abs{\braket{X_I(t)}_C -\braket{X_I(t)}_{C'}}}{\braket{X_I(t)}_C}. \label{dev_mean_pos} 
\end{equation} 
The time-evolution of $\Delta \braket{X_I(t)}_{C,C'}$ is presented in Fig. \ref{fig:14} following an interspecies interaction quench from 
$g_{BI}=0$ to $g_{BI}=2$ [Fig. \ref{fig:14} (a)] and $g_{BI}=-2$ [Fig. \ref{fig:14} (b)]. 
Evidently, a systematic convergence of $\Delta \braket{X_I(t)}_{C,C'}$ in both the repulsive and the attractive regime of interactions can be inferred. 
Focusing on repulsive interactions, we observe that $\Delta \braket{X_I(t)}_{C,C'}$ between the $C=(8;3;8)$ and $C'=(9;3;9)$ orbital configurations lies 
below $4.3\%$ for all evolution times. 
Moreover, $\Delta \braket{X_I(t)}_{C,C'}$ calculated for $C=(8;3;8)$ and $C=(7;3;7)$ is at most $5\%$ during the dynamics. 
Similar observations can be made by inspecting $\Delta \braket{X_I(t)}_{C,C'}$ in the case of a quench towards attractive interactions, see Fig. \ref{fig:14} (b). 
Indeed, the mean position when $C=(8;3;8)$ and $C'=(9;3;9)$ becomes at most of the order of $4.1\%$ while e.g. for $C=(8;3;8)$ and $C'=(6;4;6)$ it acquires 
a maximum value of $7.2\%$. 

Furthermore we showcase the convergence of the Von-Neumann entropy in the course of the time-evolution.    
More precisely, we inspect the relative difference of $S_{VN}(t)$ calculated within the $C=(8;3;8)$ and different orbital 
configurations $C'=(D;d_B;d_I)$ namely  
\begin{equation}
\Delta S_{VN}(t)_{C,C'} =\frac{\abs{S_{VN}(t)_C -S_{VN}(t)_{C'}}}{S_{VN}(t)_C}. \label{dev_Von_Neum} 
\end{equation} 
The dynamics of the relative deviation $\Delta S_{VN}(t)_{C,C'}$ is illustrated in Fig. \ref{fig:14} after an interaction quench from 
$g_{BI}=0$ to $g_{BI}=2$ [Fig. \ref{fig:14} (c)] and $g_{BI}=-2$ [Fig. \ref{fig:14} (d)] for different of orbital configurations $C'$ and fixed $C=(8;3;8)$. 
As it can be seen, convergence is achieved also for $\Delta S_{VN}(t)_{C,C'}$ at both repulsive and attractive postquench interspecies 
interaction strengths. 
For repulsive interspecies couplings, e.g. $g_{BI}=2$ presented in Fig. \ref{fig:14} (c), the deviation $\Delta S_{VN}(t)_{C,C'}$ with $C=(8;3;8)$ and $C'=(9;3;9)$ 
[$C=(7;3;7)$] is smaller than $4\%$ [$2.7\%$] throughout the evolution. 
Turning to attractive postquench interactions such as $g_{BI}=-2$ [Fig. \ref{fig:14} (d)], we deduce that $\Delta S_{VN}(t)_{C,C'}$ among the orbital configurations 
$C=(8;3;8)$ and either $C'=(9;3;9)$ or $C'=(6;4;6)$ takes a maximum value of the order of $6\%$ or $7.2\%$ respectively during the dynamics. 
We should also mention that a similar investigation has been performed for all other interspecies interaction quench amplitudes discussed in the main 
text and found to be adequately converged (not shown here for brevity).

\section*{Acknowledgements} 
S.I.M. and P.S. gratefully acknowledge financial support by the Deutsche Forschungsgemeinschaft 
(DFG) in the framework of the SFB 925 ``Light induced dynamics and control of correlated quantum
systems''. 
F.G. acknowledges support from the Technical University of Munich - Institute for Advanced Study, funded by the German Excellence Initiative 
and the European Union FP7 under grant agreement 291763, from the DFG grant No. KN 1254/1-1, and DFG TRR80 (Project F8). 
S. I. M.  gratefully acknowledges financial support by the Lenz-Ising Prize of the University of Hamburg.

{}

\end{document}